%% file: longper.tex
\documentclass{emulateapj}

\usepackage{subfigure}
\usepackage{url}
\usepackage{multirow}
\usepackage{color}
\usepackage{graphicx}
\usepackage{xspace}
\newcommand{\pc}{$P_{c}$ }
\newcommand{\mc}{$M_{c} \sin{i}_{c}$ }
\newcommand{\kai}{\ensuremath{\chi^2}}
\newcommand{\kain}{\ensuremath{\chi^2_{\nu}} }
\newcommand{\MERCURY}{{\tt MERCURY}\ }
\newcommand{\BOOTTRAN}{{\tt BOOTTRAN}\ }
\newcommand{\RVLIN}{{\tt RVLIN}\ }

\newcommand{\MPFIT}{{\tt MPFIT}\ }

\shorttitle{Updates to Long-Period Planets}
\shortauthors{Feng et al.}

\begin{document}


\title{\uppercase{The California Planet Survey IV: A Planet Orbiting the Giant Star HD 145934 and Updates to Seven Systems with Long-Period Planets \altaffilmark{*} }}


\author{\textsc{Y. Katherina Feng\altaffilmark{1,4}, Jason T.\ Wright\altaffilmark{1}, Benjamin Nelson\altaffilmark{1}, Sharon X. Wang\altaffilmark{1}, Eric B.\ Ford\altaffilmark{1}, Geoffrey W.\ Marcy\altaffilmark{2}, Howard Isaacson\altaffilmark{2}, and Andrew W.\ Howard\altaffilmark{3}}}

\altaffiltext{$*$}{Based in part on observations obtained at the W.~M.~Keck Observatory, which is operated by the University of California and the California Institute of Technology.}  

\altaffiltext{1}{Center for Exoplanets and Habitable Worlds, Department of Astronomy \& Astrophysics, 525 Davey Lab, The Pennsylvania State University, University Park, PA 16802, USA; \\Corresponding author: astrowright@gmail.com}
\altaffiltext{2}{Department of Astronomy, University of California, Berkeley, CA 94720-3411, USA}
\altaffiltext{3}{Institute for Astronomy, University of Hawaii, 2680 Woodlawn Drive, Honolulu, HI 96822, USA}
\altaffiltext{4}{Also at Department of Astronomy \& Astrophysics, 1156 High Street, MS: UCO / LICK, University of California, Santa Cruz, CA 95064, USA.}

\received{2014 August 11}
\accepted{2014 November 26}

\begin{abstract}
We present an update to seven stars with long-period planets or planetary candidates using new and archival radial velocities from Keck-HIRES and literature velocities from other telescopes. Our updated analysis better constrains orbital parameters for these planets, four of which are known multi-planet systems.  HD 24040 \textit{b} and HD 183263 \textit{c} are super-Jupiters with circular orbits and periods longer than 8 yr.  We present a previously unseen linear trend in the residuals of HD 66428 indicative on an additional planetary companion. We confirm that GJ 849 is a multi-planet system and find a good orbital solution for the \textit{c} component: it is a $1 M_{\rm Jup}$ planet in a 15 yr orbit (the longest known for a planet orbiting an M dwarf).  We update the HD 74156 double-planet system.  We also announce the detection of HD 145934 \textit{b}, a $2 M_{\rm Jup}$ planet in a 7.5 yr orbit around a giant star.  Two of our stars, HD 187123 and HD 217107, at present host the only known examples of systems comprising a hot Jupiter and a planet with a well constrained period $ > 5$ yr, and with no evidence of giant planets in between.  Our enlargement and improvement of long-period planet parameters will aid future analysis of origins, diversity, and evolution of planetary systems.  
\end{abstract}


\section{Introduction} 

\label{intro}

\subsection{Background}
The possibility of ``Earth 2.0'', and especially another planet that hosts life, drives much of the search for exoplanets. As of 2014 November, efforts over two decades have uncovered more than 1400 planets and almost 4000 planetary candidates \citep[; exoplanets.org]{eod, Burke2014}. The variety of discoveries, from lone Jupiter-mass planets in few-day orbits to packed systems with multiple planets that fit within Mercury's orbit, raises a significant question as to the nature of our solar system: Are we unique?

To search for analogs of the solar system, we target multi-planet systems and long-period giant planets, reminiscent of our own outer solar system.  Because we seek planets with orbits of at least a few hundred days, the radial velocity (RV) method of exoplanet detection is advantageous \citep[e.g.][]{wright2013, butler1996,51peg}. The RV method is the longest running, with multiple surveys studying thousands of stars. In our study, we utilize up-to-date velocities from Keck observatory's High Resolution Echelle Spectrometer \citep[HIRES;][]{vogt1994} and complementary, published velocities from other telescopes, where available.

So far, the number of planets discovered by RVs with periods greater than 1000 days is 103, only 16 of which have periods longer than 3000 days.  This is in contrast to the 336 such planets with shorter periods.\footnote{We follow \citet{eod}, who adopt a upper limit on minimum mass of 24 times the mass of Jupiter}  Our intent is to enlarge this sample of long-period planets to search for planetary systems with Jupiter analogs. Of these 103 planets, 31 are in multi-planet systems. The study of multi-planet systems addresses planetary formation, migration, and dynamics. Having a large sample can also contribute to the understanding of the evolution and lifetime of stable planetary systems. Studies can examine the orbital eccentricities and perform dynamic simulations and probe migration.

For the purposes of this discussion, we follow \citet{wang2012} and define a Jupiter analog as a planet with $P > 8$ yr, $4 > M\sin(i) > 0.5 M_{\rm Jup}$, and $e < 0.3$, but we also adopt an upper period limit of $P < 16$ yr. Of the confirmed RV planets, only 13 planets fit the above criteria \citep{eod}. Another motivation for studying systems with long-period Jupiter analogs is the role such a Jovian planet may play in the habitability of an Earth-like planet in the same system. \citet{wetherill1994} argued that Jupiter acts as a shield that deflects comets originating from the Oort Cloud or Kuiper Belt, protecting the inner solar system. Without Jupiter, \citet{wetherill1994} suggested an increase in the frequency of cometary impacts on Earth by 1000 -- 10,000 times the present--day value. Multi-planet systems serve not only as examples of planet-planet interaction but also as models for plane--comet dynamics.

\subsection{Plan}
Section \ref{method} gives an outline of the steps taken for characterizing the planetary systems. In Section \ref{results}, we describe the planetary systems orbiting seven stars.  Each of these systems already has at least one planet known and exhibits RV residuals indicative of an outer companion.  Additionally, each can have its planetary orbits significantly refined with our new velocities from Keck, and in some cases we show that an outer, decade-long planetary orbit has finally completed. We present a summary of the radial velocity data, mean uncertainties, and telescope offsets in Table \ref{rv_sum}. Table \ref{stellar} lists the stellar parameters of the target stars. Table \ref{orb} lists the orbital parameters of the planets presented in this paper. We also present figures showing the RV curves and residuals for each system. Section \ref{giant} presents an analysis of a new planet, HD 145934 \textit{b}. We discuss our findings and future prospects in Section \ref{discussion}.

\section{Methodology} \label{method}

\subsection{Radial Velocity Sources and Analysis}
We combine previously published data from other telescopes to complement the time span and quantity of Keck-HIRES observations obtained by the California Planet Survey \citep{CPS1,CPS2,CPS3} for many purposes, including as part of the $\eta_\oplus$ survey \citep{etaEarth1,etaEarth0,etaEarth2,etaEarth3}. At the time of the first confirmed RV planet, 51 Pegasi \textit{b} \citep{51peg}, several surveys were underway and actively monitoring stars for the signs of planets \citep[e.g.][]{cochran1994,fischer2014}. The discovery team for 51 Pegasi used the ELODIE spectrograph \citep{ELODIE}, which was part of the Northern Extrasolar Planet Search until the SOPHIE spectrograph \citep{SOPHIE} replaced it in 2006. The CORALIE spectrograph \citep[e.g.~][]{CORALIE} was situated in Chile as part of the Southern Sky extrasolar Planet search Programme.  It has been joined by HARPS \citep{HARPS}, also located in Chile.  We make use of literature data from all four spectrographs in this work.  Other data come from High Resolution Spectrograph (HRS) of the Hobby-Eberly Telescope \citep{HRS}, the Tull Spectrograph at the 2.7-m telescope of McDonald Observatory \citep{Tull}, and the Hamilton spectrograph at Lick Observatory \citep{Hamilton}.

To analyze and fit the data, we use the \citet{wright2009} \RVLIN package written in IDL that naturally handles multiplanet systems using data from multiple telescopes in systems where planet--planet interactions are negligible given the precision and the span of the observations. In this package, RV curves are described by both non-linear and linear parameters, and the package performs least-squares fitting on them separately. The package uses a simple linear least-squares solution for the linear parameters, and the Levenberg--Marquardt algorithm for the nonlinear parameters. \RVLIN supplies a sum-of-Keplerians model (plus optional secular trend and offsets between instruments) to \MPFIT, the IDL implementation of the LM method developed by \citet{markwardt}.

%For a system with $n$ planets, its RV curve needs to be fit with $5n+1$ parameters. Each planet has a semiamplitude to the RV curve, $K$ (m s$^{-1}$); a period, $P$ (days); an eccentricity to its orbit, $e$, where $e=0$ is circular; an argument of periastron, $\omega$ (an angle), where the planet passes closest to the star; a time of periastron passage, $T_{\rm p}$; and an apparent RV of the center of mass of the system, $\gamma$. RVLIN provides an option of including a value for trend, $dv/dt$. Equation \ref{msini} can be used to solve for the mass of the planet, $m$ and the mass function of a system, shown in the large parentheses:
%\begin{equation}
%K^3 = \frac{2 \pi G}{P(1-e^2)^{{\frac{3}{2}}}}\left(\frac{m^3 \sin^3 i}{(M_{\star}+m)^2}\right),
%\label{msini}
%\end{equation}
%where $M_{\star}$ is the mass of the host star, $i$ is the inclination of the orbit with respect to the plane of the sky ($90^{\circ}$ is edge-on), and $G$ is Newton's gravitational constant.

We fit for these new planetary system as follows: (1) we collect Keck RVs before and after the 2004 HIRES upgrade separately to account for any (small) offsets between the pre- and post-upgrade time series \citep[e.g.,][]{kane2014};  (2) we collect literature RVs for the system from other telescopes, if available (3) we use the published orbital parameters \citep[which we collect from the Exoplanet Orbit Database;][]{eod} as initial guesses for the planets' orbits; (3) we use \RVLIN to fit the system anew (with additional planets contributing five model parameters each, if necessary); (4) we use the reduced \kai ($\kai_\nu$) to describe the goodness of fit.

We calculate most orbital parameter uncertainties using \BOOTTRAN \citep{wang2012}, which uses \RVLIN and a bootstrapping method to compute the distribution of parameters consistent with the data. Because uncertainties can be highly non-Gaussian for planets with incomplete orbits, we also examine the minimum \kai\ surface in minimum mass-period space (Section~\ref{results}).

For our fits, we choose values for the jitter  \citep[][and references therein]{jitter} that yield $\kai_\nu$ values close to 1; usually we pick a value similar to the rms of the initial fit which does not incorporate jitter. If a star has data from several (more than three) instruments taken by multiple teams, we apply jitter on an instrument-by-instrument basis. To do so, we ran the fit with no assumed jitter, calculated for each instrument the standard deviation of the residuals, and added that value in quadrature to the velocities. After that, we rerun the fit and that yielded the best-fit parameters. We utilized an instrument-by-instrument jitter for HD 24040, HD 74156, and HD 217107. In general, we are confident in the relative instrumental uncertainties in the pre- and post-upgrade HIRES data, and we use a common jitter value for both.\footnote{The effect of jitter on the best fit values of an orbital solution is to give more even weight to points with different measurement uncertainties; in the cases of the well-detected planets we discuss in this work, the exact value of the jitter has very little effect on these best-fit values.  Because we determine most of our parameter uncertainties via bootstrapping, our uncertainties are not strongly affected by our choice of jitter, and so there is no need to find the precise jitter value that yields $\kai_\nu = 1.0$.} Table \ref{stellar} lists the stellar parameters of host stars, and Table \ref{orb} lists the orbital parameters of the planets we discussed below. Corresponding RV plots and additional figures follow the text.

Sun-like stars are known to have magnetic cycles with periods comparable to the period of Jupiter \citep{baliunas1995}.  A persistent concern in the hunt for long-term RV signals from Jupiter analogs has been that they might be mimicked by the effects of such magnetic cycles, which could alter convective patterns such that the magnitude of the disk-integrated convective blueshift of a star might vary with the stellar cycle \citep{dravins1985, walker1995, deming1987,santos2010}.  

A common way to check that magnetic effects are not responsible for RV variations is to measure correlations between the RVs and activity indices such as \ion{Ca}{2} H \& K.  Previous work by \citet{wright2008}, \citet{santos2010}, and \citet{lovis2011} find that the observed activity-cycle-induced RV amplitudes are typically quite small (a few m s$^{-1}$ or less), although there are suggestions that a few stars may show abnormally high levels of correlation (at the level of 10--20 m s$^{-1}$).

We have checked for activity cycles in these stars to see if they have similar periods and phases to the RV measurements.  To do this, we have used the \ion{Ca}{2} H \& K chromospheric activity measurements from \citet[hereafter W04]{wright2004}, \citet[hereafter IF10]{isaacson2010}, and more recent measurements made using the same data stream and pipeline as the latter work.  For some stars, there appear to be calibration differences between the measurements published by W04 and those made using the IF10 pipeline, necessitating a rescaling or application of an offset to one of the streams.  This is most apparent in ``flat activity" stars which show no variation but occasionally exhibit a large jump in activity level between the two data streams.

In six of our stars, there is no appreciable activity variation \citep[i.e., they are ``flat activity'' stars, ][]{saar1998}, making it very unlikely that the large RV variations we see are due to solar-type activity cycles.  The seventh star, HD 183263 does show a significant cycle, however.  The W04 activity levels decrease from 2002 to 2004, and the IF10 show a continued decrease starting in late 2004, which bottoms out in a minimum in 2012.  The actual velocities show a minimum in 2005 and a maximum in 2012, thus exhibiting a shorter period than the actual activity cycle.  The negative correlation between 2005--2012 and the very high amplitude of the RV signals are inconsistent with typical stars with RV-activity correlation seen in \citet{wright2008} and described by \citet{lovis2011}.  It is thus very unlikely that any of the long-period signals we describe in this work are due to stellar magnetic activity cycles.

\subsection{Minimum Masses from Linear Trends Alone}

 In some cases, we find that a secular increase or decrease in the observed radial velocities is present (a ``linear trend''), which is presumably a small portion of a Keplerian signal from a massive companion, typically an outer planet, or a secondary star or brown dwarf \citep{TRENDS1,TRENDS2,TRENDS3,TRENDS5,montet2014,Knutson14}.

\begin{figure}
\includegraphics[width=0.4\textwidth]{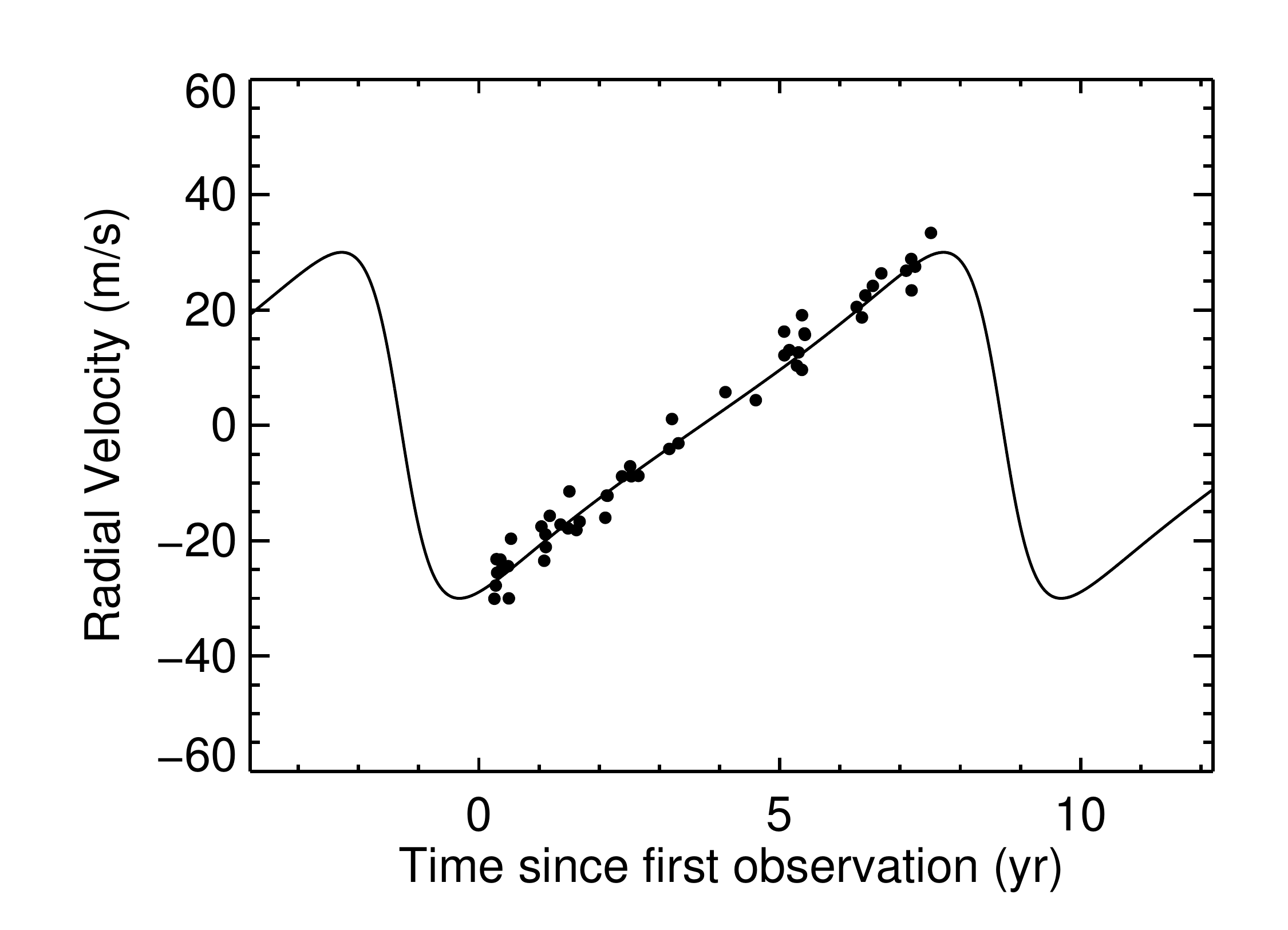}
\includegraphics[width=0.4\textwidth]{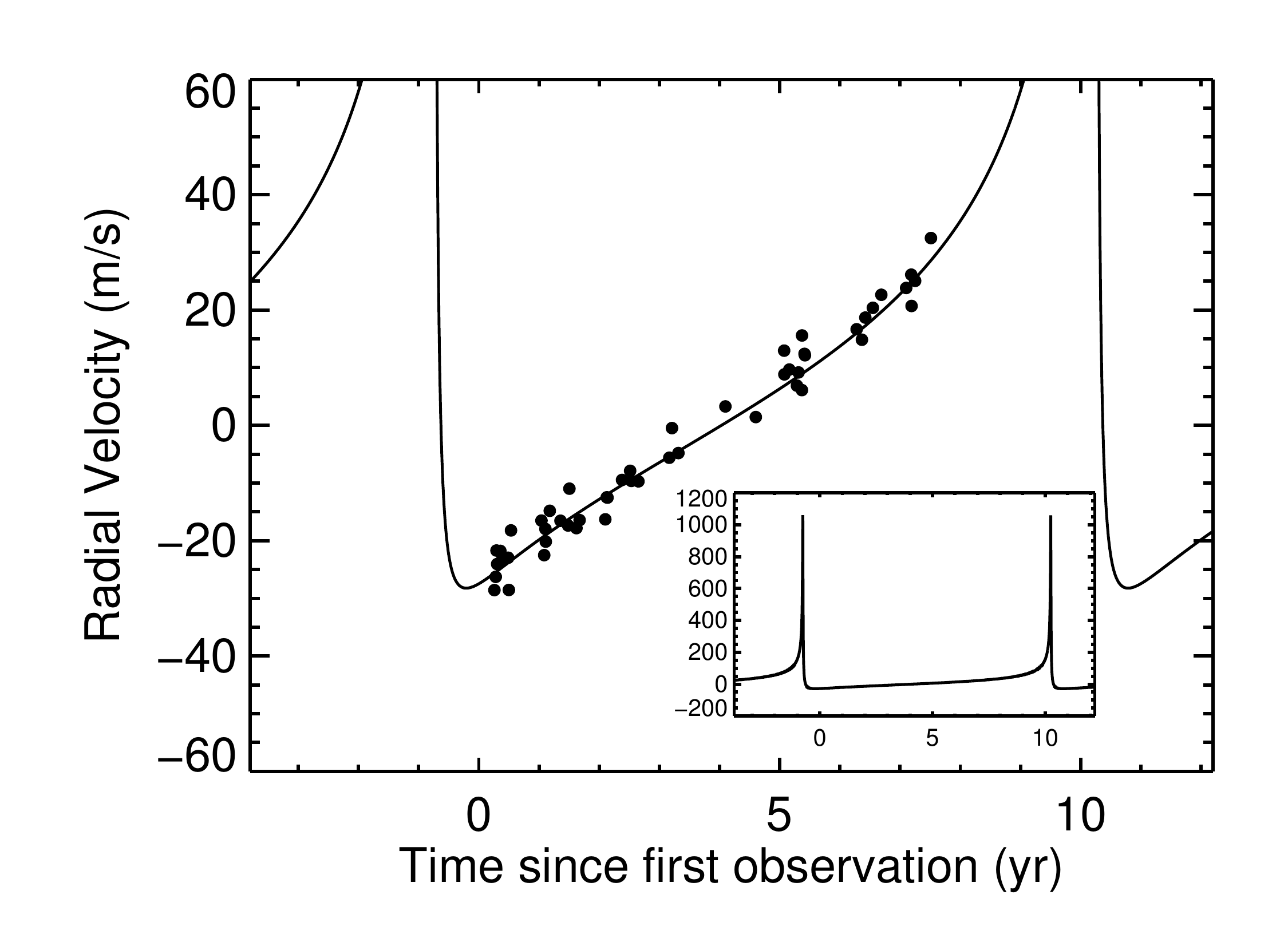}
\includegraphics[width=0.4\textwidth]{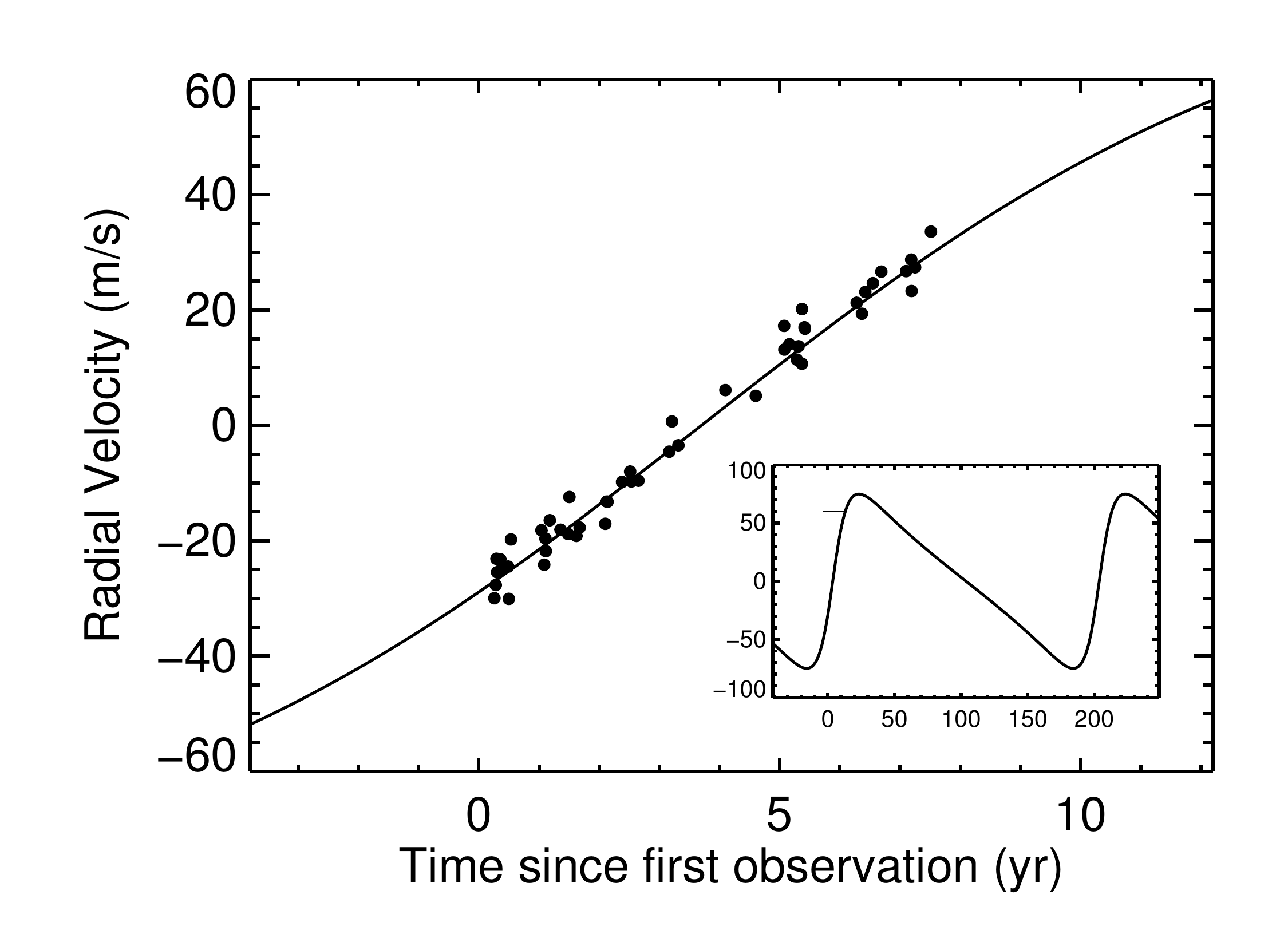}
\caption{Fifty synthetic RV measurements made over 8 yr by an unlucky observer of a hypothetical system with Gaussian errors of 3 m s$^{-1}$.
{\it Top:} RV curve of a planet with $M \sin i = 1.97 M_{\rm Jup}$, $P=10$ yr, $K=30$ m s$^{-1}$, $e=0.5$ and
  $\omega = 90^{\arcdeg}$.  The observer might conclude, incorrectly, that they were seeing the effects of a
  distant exoplanet with $P \gg 8$ yr  and $K \gg 30$ m s$^{-1}$.  
{\it Middle:} RV curve of a planet with $M \sin i = 6.6 M_{\rm Jup}$, $P=11$ yr, $K=345$ m s$^{-1}$,
    $e=0.97$ and $\omega = 20^{\arcdeg}$. Although the
    period $P$ and magnitude of the observed trend are about the
    same as that in the top panel, the true semi-amplitude of the orbit is much
    larger.  The inset illustrates the complete RV curve, with
    the same units as the main figure. 
{\it Bottom:} RV curve for hypothetical
    planet with $M \sin i = 11.7 M_{\rm Jup}$, $P=200$ yr, $K=65$ m s$^{-1}$, 
    $e=0.5$ and $\omega = 270^{\arcdeg}$. Although the magnitude of the
    observed trend is the same as that in the top and middle panels, the period in this case is much longer, while $K$
    is only modestly larger.  The inset illustrates the RV curve, with
    the same units as the main figure, over more than a complete
    orbit.  The box in the inset illustrates the span of the main panel.
All three panels are reproduced from \citet{wrightthesis}. \label{fooled1}}
\end{figure}

When the trend shows no curvature and we have no AO imagery to put limits on the mass and angular separation of companions, we usually say very little about the companion beyond a minimum mass (and a maximum luminosity from the fact that its spectrum did not complicate the RV analysis).  The scenario that gives the minimum mass to a companion generating a linear trend of a given magnitude is one that has $e\sim 0.5$, $\omega=90^{\arcdeg}$, which produces a sawtooth-like RV curve with a long, nearly linear component for $\sim 80\%$ of the orbit with a brief, high-acceleration component during periastron for the other $\sim 20\%$ (\citet{wrightthesis}, and see top panel of Figure \ref{fooled1} for an example; other panels show other pathological cases with radically different periods and semiamplitudes that mimic the same trend).  

The minimum mass of a planetary companion detected only by its strongly detected constant acceleration $\dot{\gamma}$, is thus derived by solving the mass function for the minimum mass \citep[e.g.,][]{wright2009} assuming $P\sim 1.25 \tau$ (where $\tau$ is the span of the observations), $e \sim 0.5$, and $K \sim \tau \dot{\gamma}$:
\begin{equation}
\label{mm}
M_{\rm minimum} \approx (0.0164 M_{\rm Jup})  \left(\frac{\tau}{\rm yr}\right)^{4/3} \left|\frac{\dot{\gamma}}{{\rm m s}^{-1}{\rm yr}^{-1}}\right| \left(\frac{M_*}{M_{\odot}}\right)^{2/3}.
\end{equation}

\section{Refined Orbital Parameters for Seven Planetary Systems} \label{results}

Our sample includes many known planetary systems of interest because of the presence of a linear trend in the residuals indicative of an additional companion; some with known trends with significant curvature; and some known to have outer companions with poorly-constrained parameters.  The first six, HD 24040, HD 66428, HD 74156, HD 183263, HD 187123, and HD 217107 are G stars; the seventh GJ 849, is an M dwarf. 

Table \ref{rv_sum} presents the time span of sets of observations, the number of points from each set, the number of new points, the mean uncertainty in velocities from each set, and the offsets between instruments. 

%We find a trend in the HD 66428 system has not been previously reported. HD 24040 \textit{b} and HD 187123 \textit{c} are Jupiter analogs. We confirm the existence of GJ 849 \textit{c}, and find that its 15-year orbital period is the longest of known planets around M dwarfs. 

\subsection{HD 24040}

\label{24040}
\citet{wright2007} reported a substellar companion to the star HD 24040 with a wide range of possible periods ($10$ yr $< P < 100$ yr) and minimum masses ($5 < M \sin i < 20 M_{\rm Jup}$). \citet{boisse2012}, combining velocities from HIRES, SOPHIE, and ELODIE, determined an orbit of $3668^{+169}_{-171}$ days (corresponding to ~10 yr) and a minimum mass of $4.01 \pm 0.49 M_{\rm Jup}$ for HD 24040 \textit{b}.  \citet{boisse2012} also found a linear trend of $3.85^{+1.43}_{-1.29}$ m s$^{-1}$ yr$^{-1}$, indicative of a third body in the system. \citet{boisse2012} also investigated potential long-term correlation between SOPHIE measurements and stellar activity indices but did not find such behavior. 

\begin{figure}
\centering
\subfigure[HD 24040 \textit{b}]{
    \includegraphics[width=0.4\textwidth]{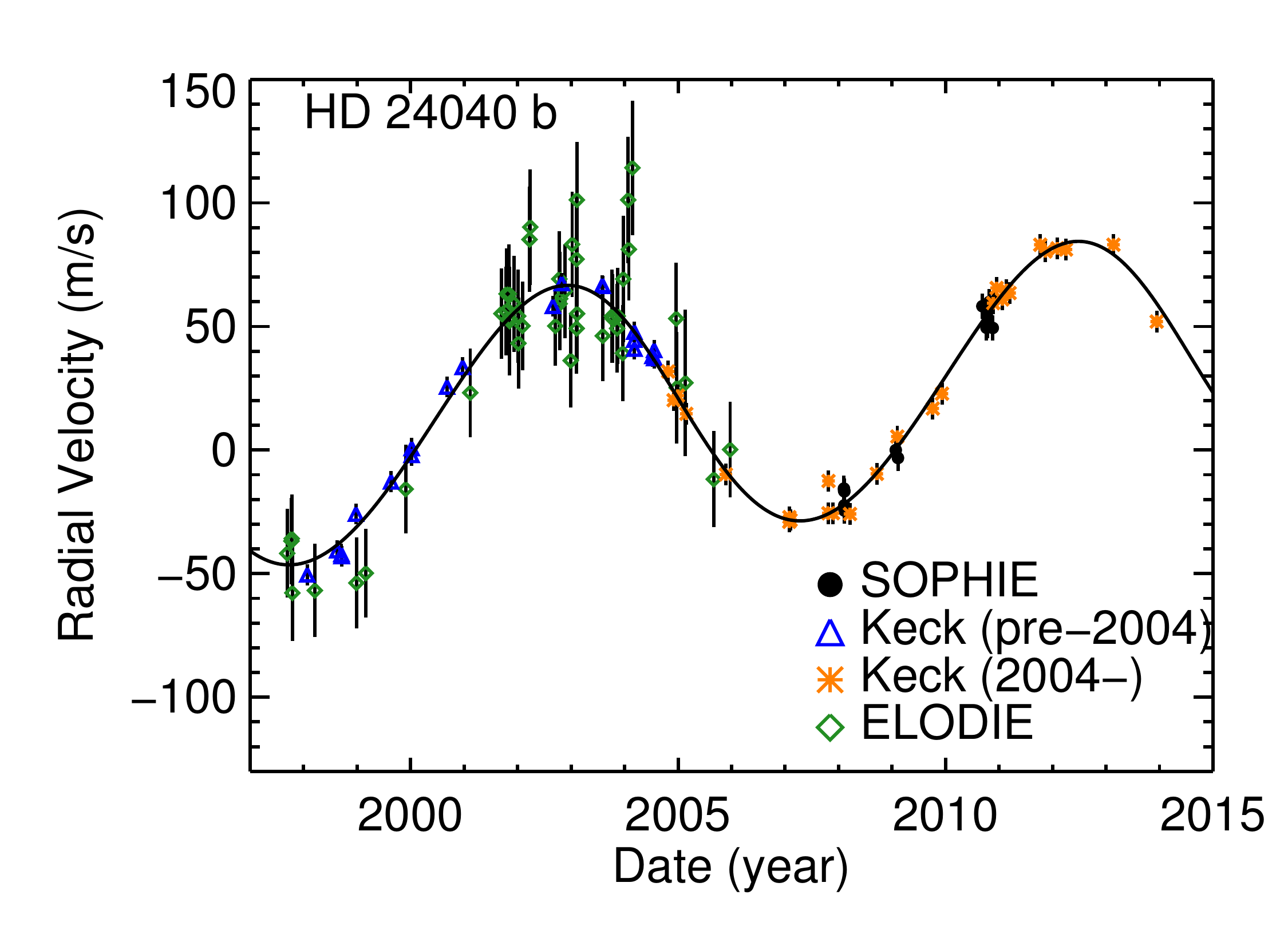}
    \label{24040sys}}
\quad
\subfigure[Residuals]{
    \includegraphics[width=0.4\textwidth]{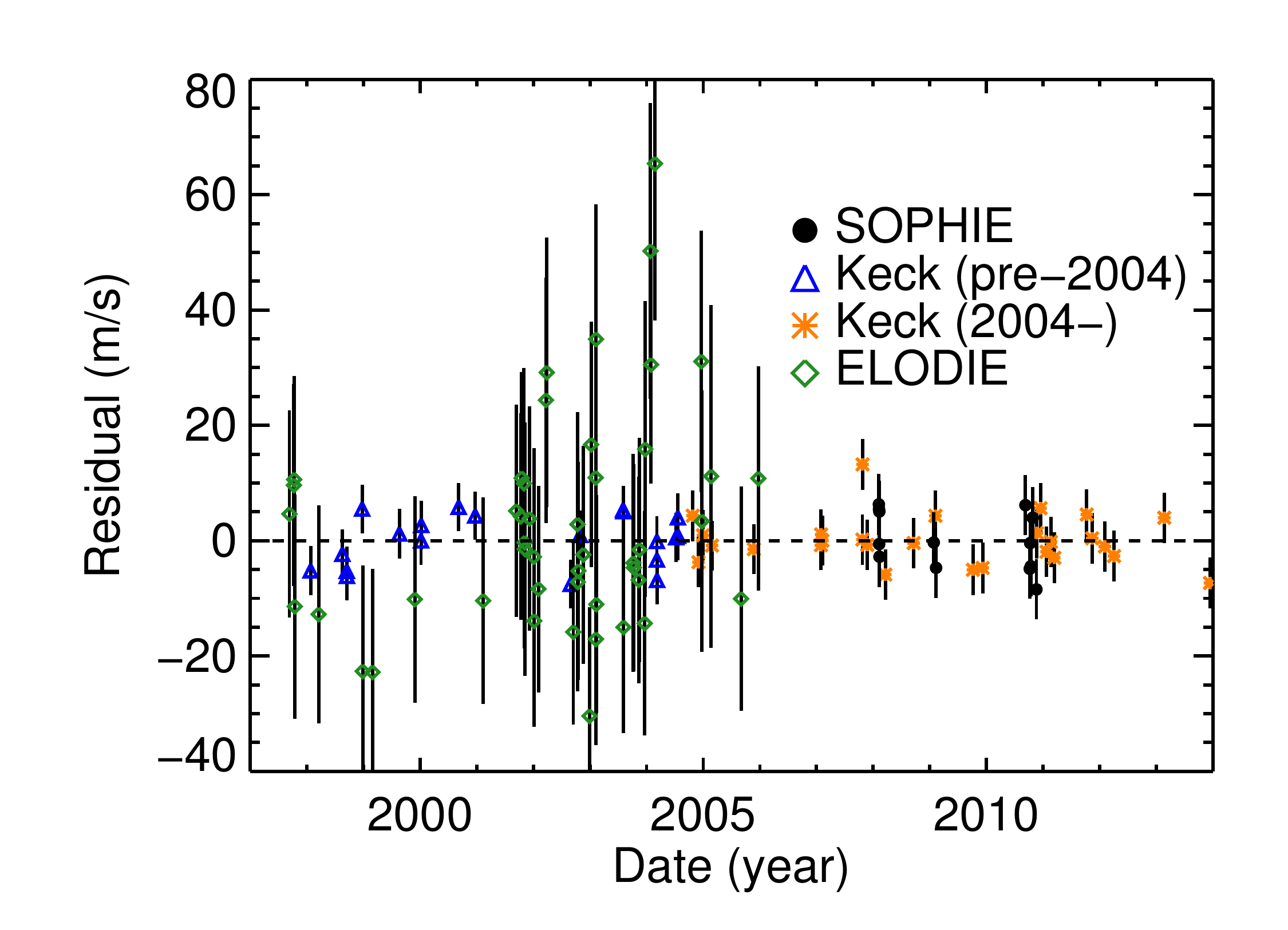}
    \label{24040res}}
\caption[Radial velocity and Keplerian fit for HD 24040 \textit{b}.]{Radial velocity and Keplerian fit for HD 24040 \textit{b}. Solid lines represent the best-fit Keplerian orbits. The fit includes a linear trend of 1.8 $\pm$ 0.4 m s$^{-1}$ yr$^{-1}$.

\ref{24040sys} Keck, SOPHIE, and ELODIE RVs overplotted by best-fit one-planet Keplerian model. \ref{24040res} Residuals of the RVs with the best-fit one-planet Keplerian model subtracted.}
\label{24040plot}
\end{figure}

 We present an updated fit with more recent Keck-HIRES velocities, seen in Figure \ref{24040plot} and Table \ref{24040rv}. We use HIRES data and published SOPHIE and ELODIE data, so in our fit we applied jitter instrument-by-instrument. With 107 velocities in total, 47 of which are from HIRES, 13 from SOPHIE, and from 47 ELODIE \citep{boisse2012}, we find for the best-fit one-planet Keplerian model an rms of 13.62 m s$^{-1}$ and \kain of 0.93. HD 24040 \textit{b} orbits at a semimajor axis of $4.637 \pm 0.067$ AU, corresponding to a period of 9.5 yr, making it a good Jupiter analog in terms of its orbit (however, its minimum mass is $4.10 \pm 0.12 M_{\rm Jup}$).  The linear trend is $1.8 \pm 0.4$ m s$^{-1}$ yr$^{-1}$ \citep[lower than reported in][]{boisse2012}, a minimum mass of at least 1.44 $M_{\rm Jup}$ according to Equation~(\ref{mm}). Our fit for HD 24040 \textit{b}, with a period of $3490 \pm 25$ days and minimum mass of $4.10 \pm 0.12 M_{\rm Jup}$, is in good agreement with the solution from Boisse et al. (2012).

\subsection{HD 66428}

\begin{figure}
\centering
\subfigure[HD 66428 \textit{b}]{
    \includegraphics[width=0.4\textwidth]{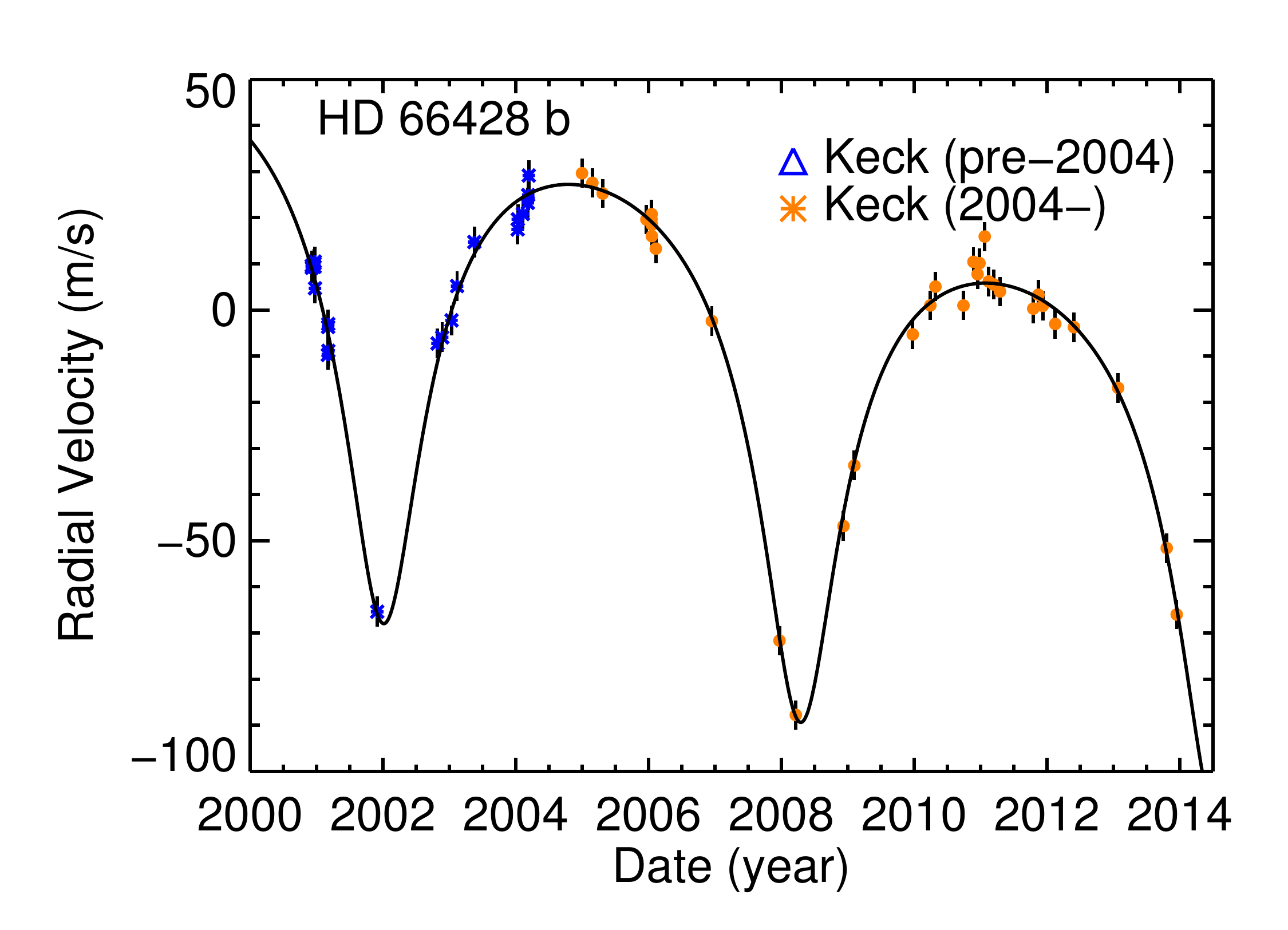}
    \label{66428sys}}
\quad
\subfigure[Residuals]{
    \includegraphics[width=0.4\textwidth]{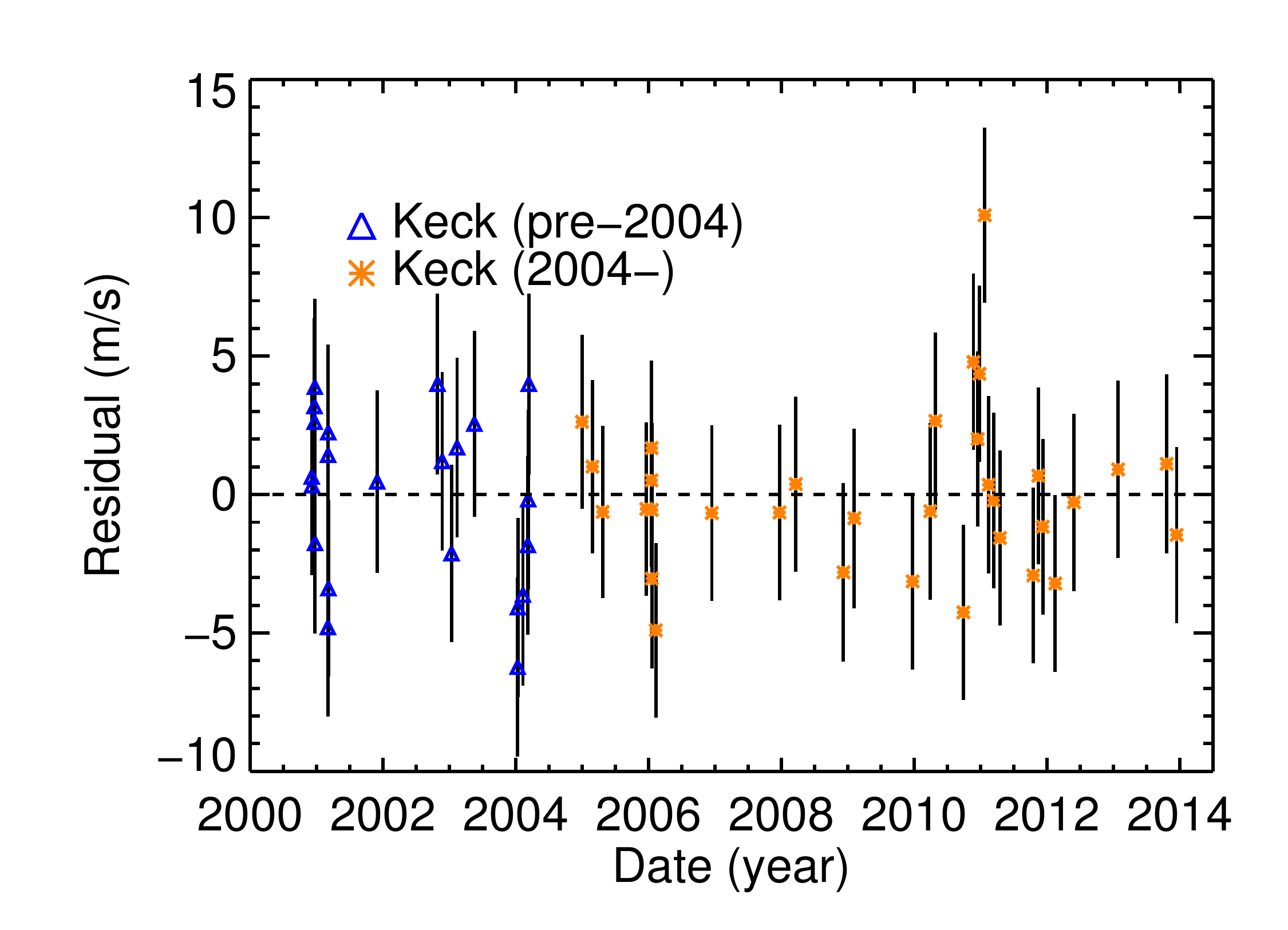}
    \label{66428res}}
\caption[Radial velocity and Keplerian fit for HD 66428 \textit{b}.]{Radial velocity and Keplerian fit for HD 66428 \textit{b}, with a trend of 3.4 m s$^{-1}$ yr$^{-1}$ incorporated. Solid lines represent the best-fit Keplerian orbits.
\ref{66428sys} Keck RVs overplotted by best-fit one-planet Keplerian model. \ref{66428res} Residuals of the RVs with the best-fit one-planet Keplerian model subtracted.}
\label{66428plot}
\end{figure}

\citet{butler2006} announced HD 66428 \textit{b}, a planet with $P = 1973 \pm 31$ d (5.4 yr), $e = 0.465 \pm 0.030$, and $M\sin(i) = 2.82 \pm 0.27 M_{\rm Jup}$. We update the orbital parameters with a total of 55 velocities from HIRES (see Figure~\ref{66428plot}). The original fit used 29 velocities taken with HIRES from 2000 to 2006. Our new fit adds 26 new data points through late 2013. Capturing two complete orbits of HD 66428 \textit{b}, the fit has an rms of 3.14 m s$^{-1}$ where we assumed a jitter of 3 m s$^{-1}$ and \kain of 0.96.  We determine a period of $2293.9 \pm 6.4$ days, or 6.3 yr. We determine a minimum mass of $3.195 \pm 0.066 M_{\rm Jup}$, which is more massive than reported in \citet{butler2006}. 

Given our larger set of radial velocities, it is understandable that our solution does not match with the solution announced in \citet{butler2006}. The final fit finds a previously unreported linear trend of $-3.4$ $\pm$ 0.2 m s$^{-1}$ yr$^{-1}$ (corresponding to a minimum mass for the outer companion of at least 1.77 $M_{\rm Jup}$, by Equation~(\ref{mm})). 

We run the fit with no jitter and no trend in order to see the significance of the detected trend. For that case, \kain is 52.56, and the rms of the residuals is 7.46 m s$^{-1}$. To compare, we found an rms of 3.14 m s$^{-1}$ and \kain of 8.23 for seven free parameters (including the trend) and no jitter. Given the improvement in the fit with a trend included, the trend is significant. We also note that the eccentricity of the orbit is large: $0.442 \pm 0.016$. The trend may indicate that the outer companion has influenced the orbit of the \textit{b} component. Further monitoring will determine the nature of the source of the trend (i.e., whether it is due to a stellar or planetary companion).

\subsection{HD 74156}

\citet{naef2004} described the HD 74156 two-planet system as a 1.86 $\pm 0.03 M_{\rm Jup}$ planet in a 51.64 $\pm$ 0.011 day period with a 6.17 $ \pm 0.23M_{\rm Jup}$ outer companion in a 5.5 yr orbit.  Multiple authors have suspected a third planet in the system. \citet{barnes2004} predicted one based on the Packed Planetary System hypothesis, and \citet{bean2008} claimed the discovery of a companion with $P = 336$ days as the third planet. Based on analysis of RV jitter, \citet{baluev2009} questioned the validity of HD 74156 ``\textit{d}'' as a false detection due to annual systematic errors from HRS. \citet{wittenmyer2009} concluded that the third planet was unlikely to be real, and \citet{meschiari2011} updated the system with further observations and reached the same conclusion.

\begin{figure}
\centering
\subfigure[HD 74156 system]{
    \includegraphics[scale=0.32]{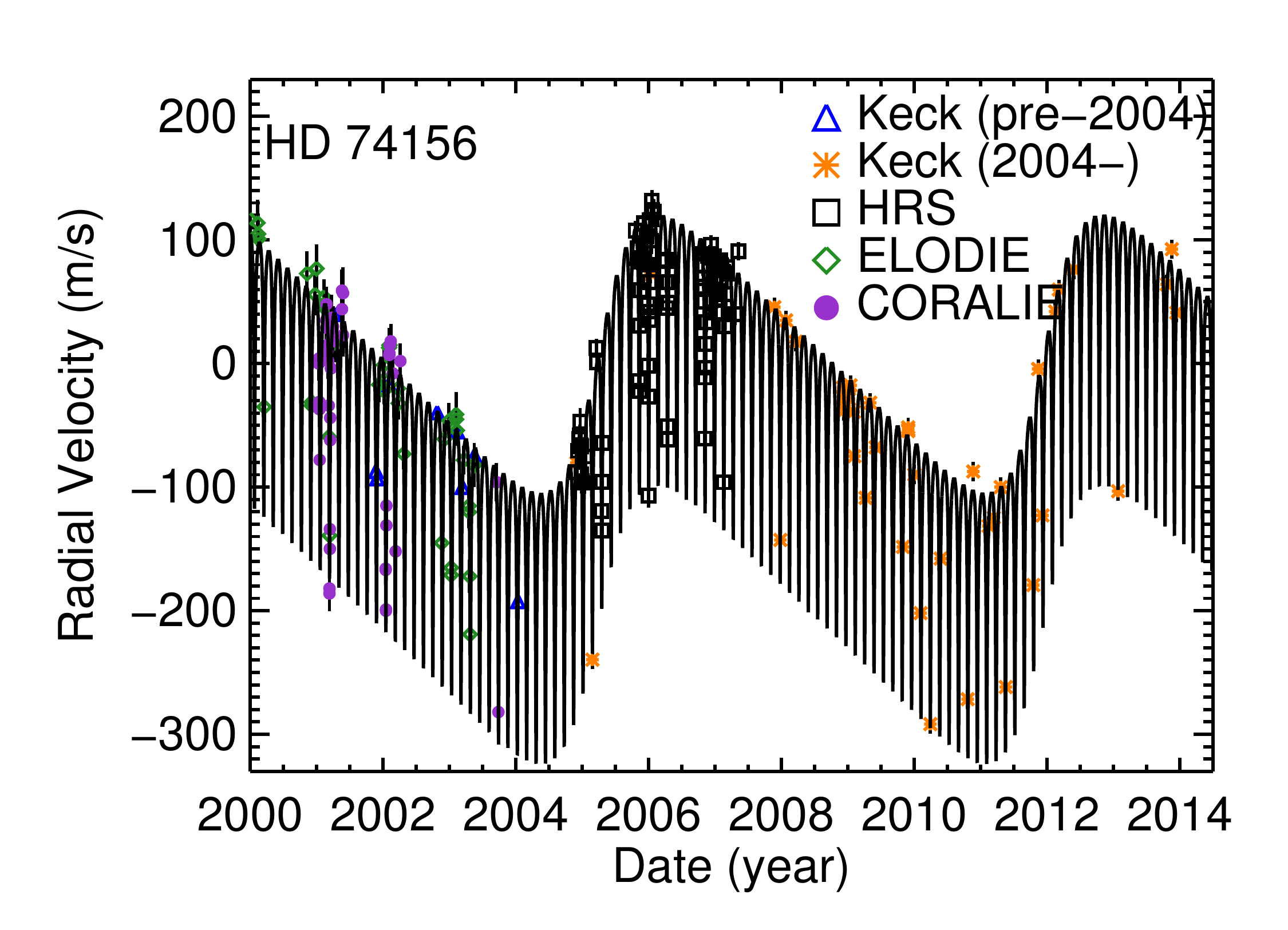}
    \label{74156sys}}
\quad
\subfigure[Residuals]{
    \includegraphics[scale=0.32]{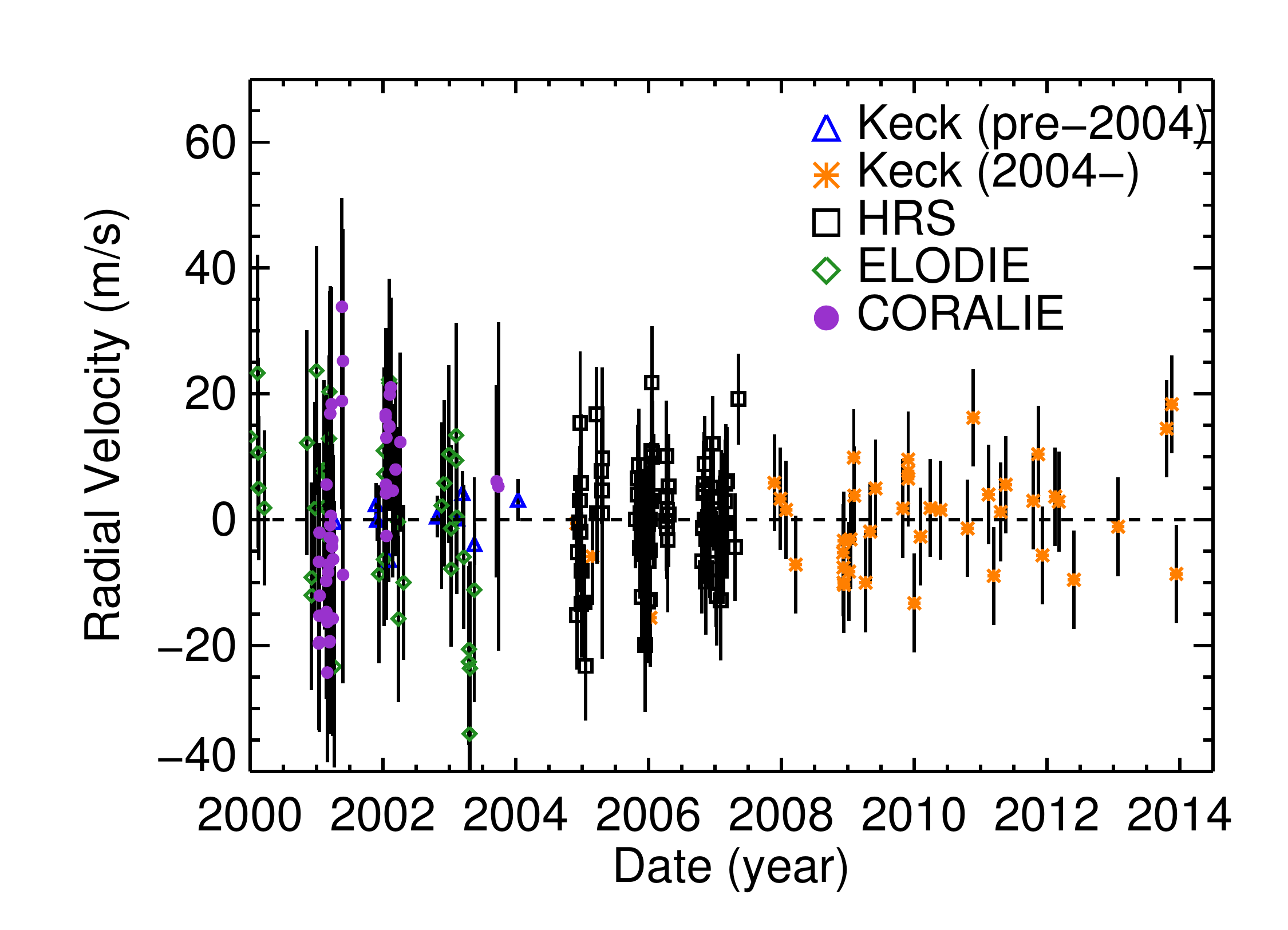}
    \label{74156res}}
\subfigure[HD 74156 \textit{b}]{
    \includegraphics[scale=0.32]{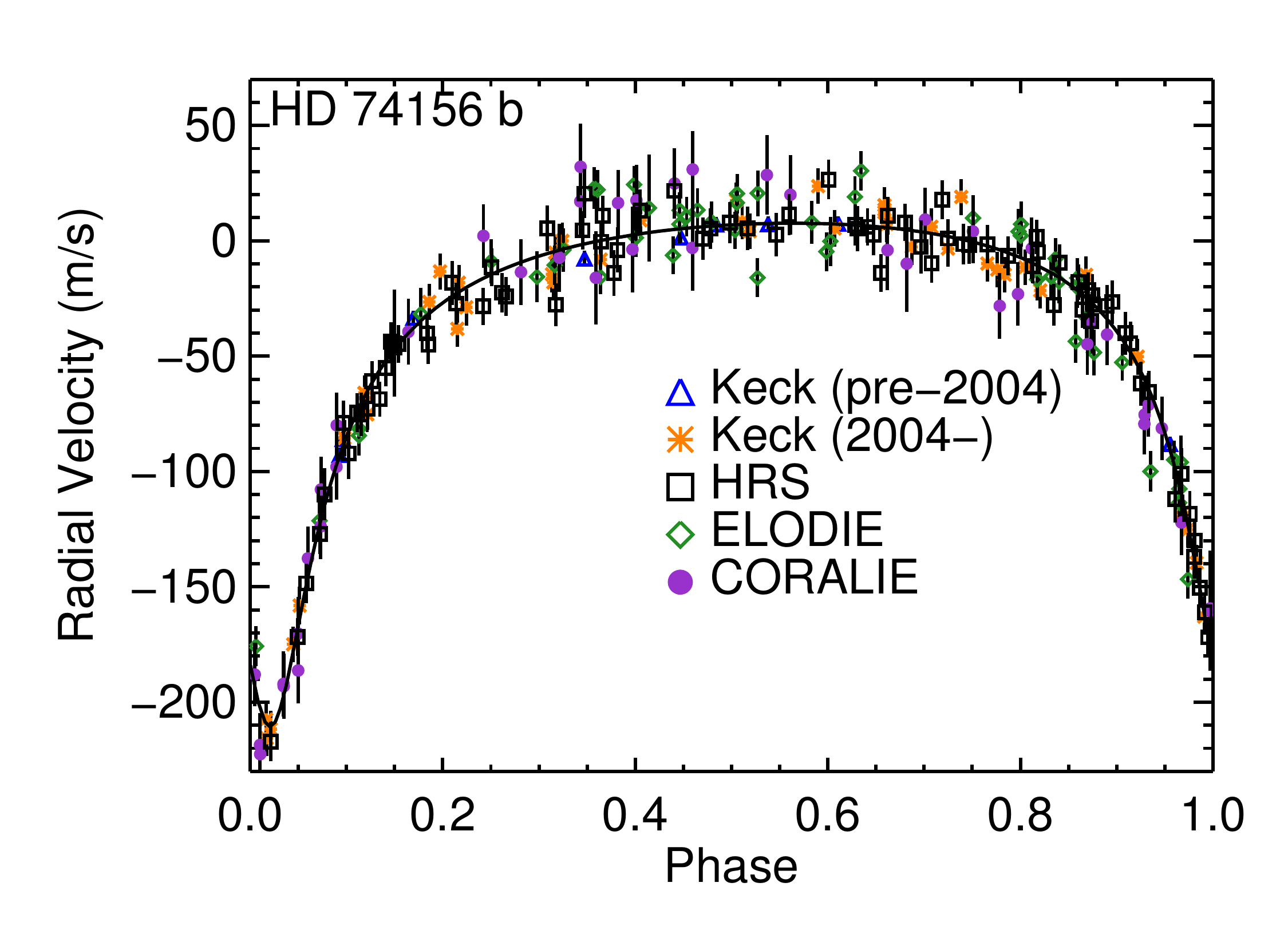}
    \label{74156b}}
\quad
\subfigure[HD 74156 \textit{c}]{
    \includegraphics[scale=0.32]{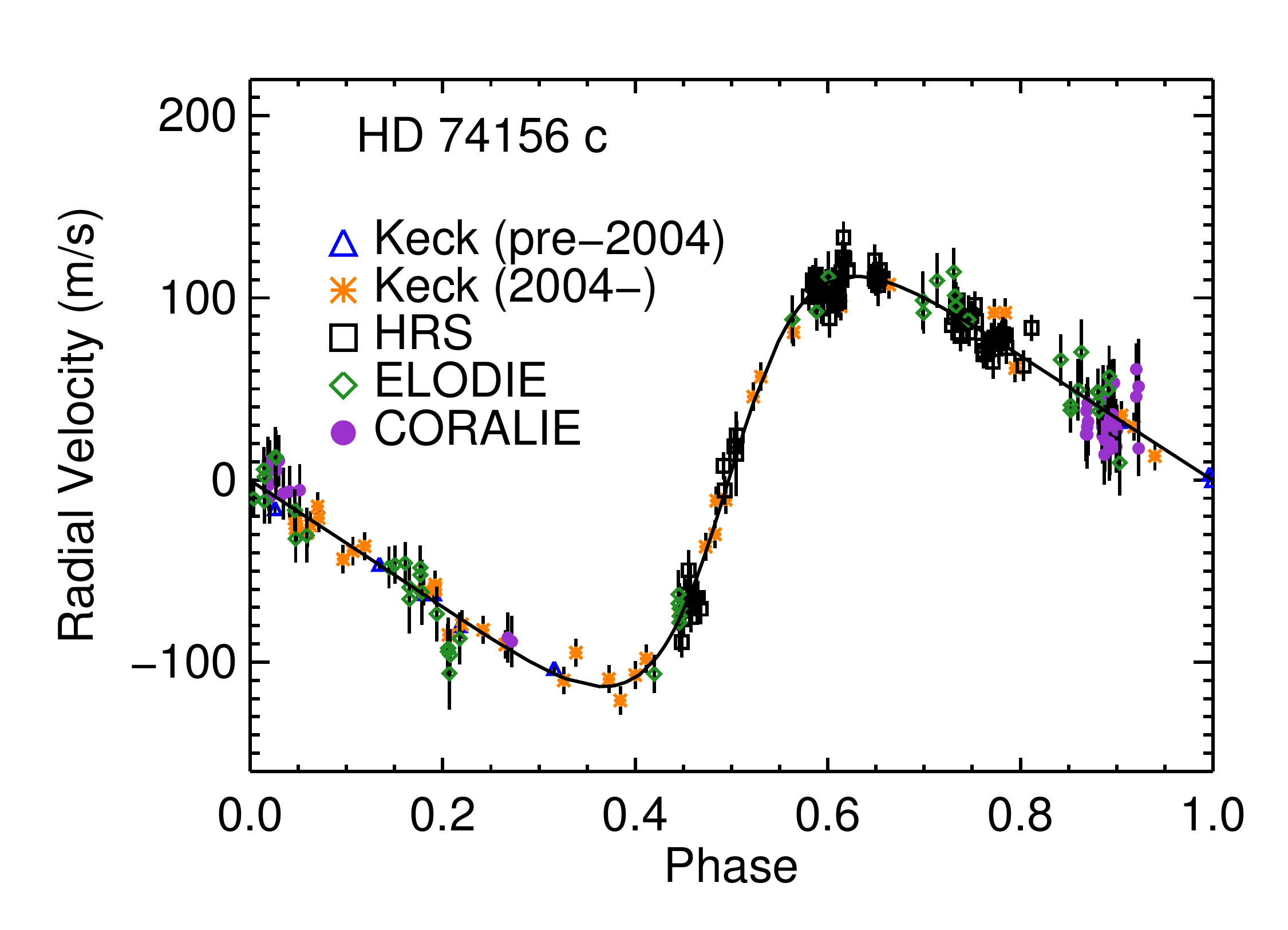}
    \label{74156c}}
\caption[Radial velocity and Keplerian fits for the HD 74156 system.]{Radial velocity and Keplerian fits for the HD 74156 system. Solid lines represent the best-fit Keplerian orbits. 

\ref{74156sys} CORALIE, ELODIE, HRS, and Keck RVs overplotted by best-fit two-planet Keplerian model.  \ref{74156res} Residuals of the RVs with the best-fit two-planet Keplerian model subtracted. \ref{74156b} and \ref{74156c}: the RV curves for HD 74156 \textit{b} and \textit{c}, respectively.}
\label{74156plot}
\end{figure}

Here, we combine 226 velocities from CORALIE and ELODIE \citep[44 and 51 observations][]{naef2004}, HRS \citep[82][]{bean2008}, and HIRES (52) (see Figure~\ref{74156plot}). We apply a two-planet Keplerian model. We added jitter instrument-by-instrument, and our fit has an rms of 11.03 m s$^{-1}$ and \kain of 0.97. We have captured at least two orbits of HD 74156 \textit{c}, making our orbital solution more robust than previously reported solutions. Table \ref{orb} lists the orbital parameters. HD 74156 \textit{c} is one of the more massive planets we have examined, with minimum mass $7.997 \pm 0.095 M_{\rm Jup}$. Both planets have large orbital eccentricities ($e = 0.64$ and $e = 0.38$ for \textit{b} and \textit{c} respectively).

\begin{figure}
\plotone{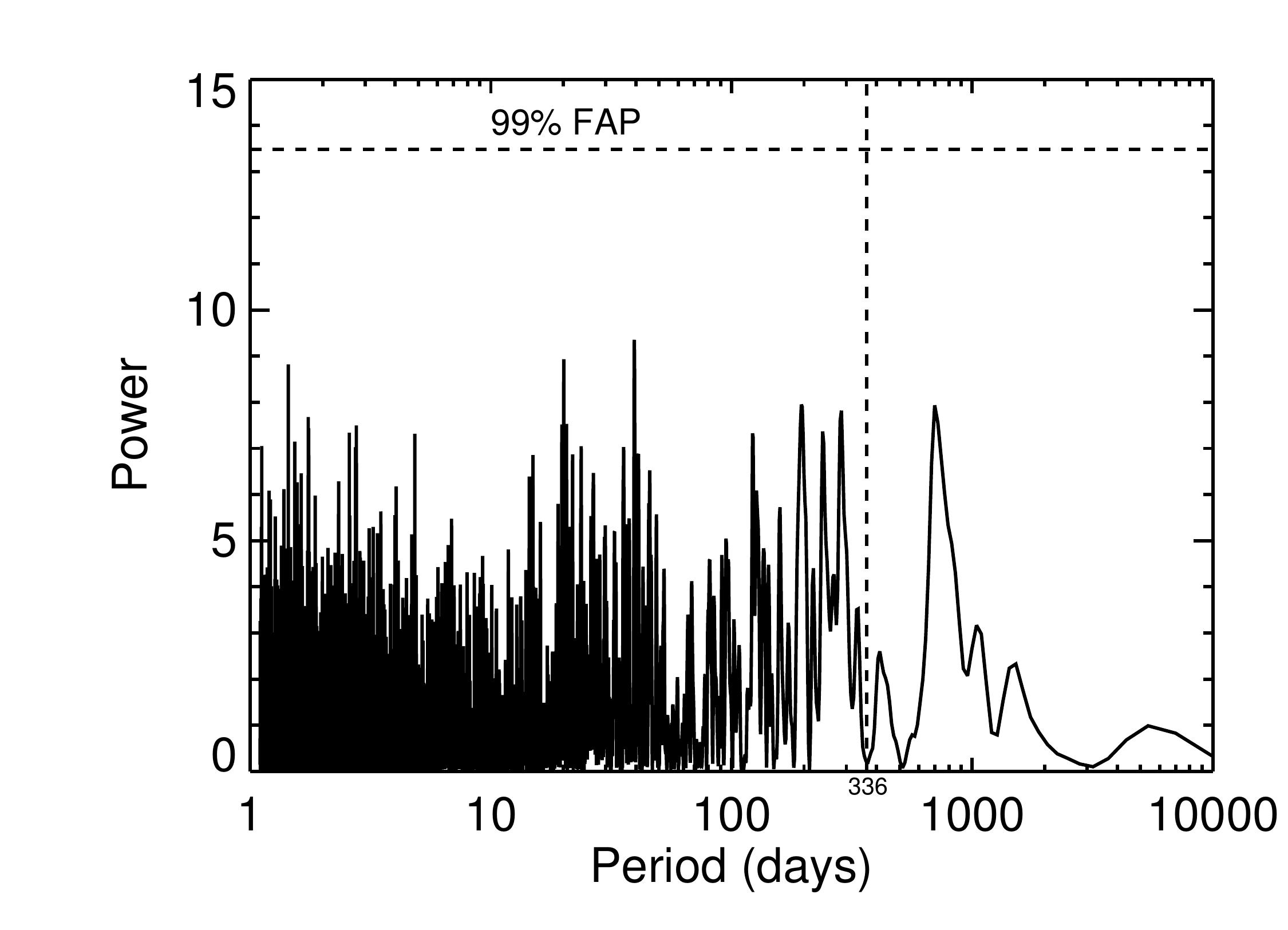}
\caption{Periodogram of the residuals to our best two-planet, five-instrument fit to the RV data for HD 74156.  There is no indication of significant power at any period, or of any power at all at 336 days, the period of the purported but disproven $d$ component.  We have computed the 99\% false alarm probability in this figure by calculating the highest peak in each of 10,000 such periodograms calculated for synthetic data sets of RV residuals \citep[e.g.,][]{etaEarth1}.  We calculated each of the 10,000 synthetic sets by randomly assigning the actual residuals (drawn with replacement) to each of the times of observations of the actual observations.  In 99\% of cases the tallest peak had power below 13.5. \label{pergram}}
\end{figure}

In Figure~\ref{pergram} we plot the Lomb--Scargle periodogram \citep{Scargle82,Horne86} of the residuals to our best two-planet fit.  There is no indication of any power at the period of the purported $d$ component, a result which is consistent with prior refutations of this signal (indeed, our analysis here uses much of the same data as previous work on the topic).  Indeed, there is no hint of significant power at any period, indicating that there is no detectable third planetary companion in this system.

\subsection{HD 183263}

First reported by \citet{marcy2005}, the HD 183263 system showed a residual linear trend in addition to a 3.7$M_{\rm Jup}$ planet in a 634-day period. \citet{wright2007} attributed the new and significant curvature in the residuals to an outer companion. \citet{2009wright} followed up and constrained the minimum mass (3.57 $\pm$ 0.55 $M_{\rm Jup}$) and period (8.4 $\pm$ 0.3 yr) for the outer companion, HD 183263 \textit{c}, to which we report an updated set of parameters.

\begin{figure}
\centering
\subfigure[HD 183263 system]{
    \includegraphics[scale=0.32]{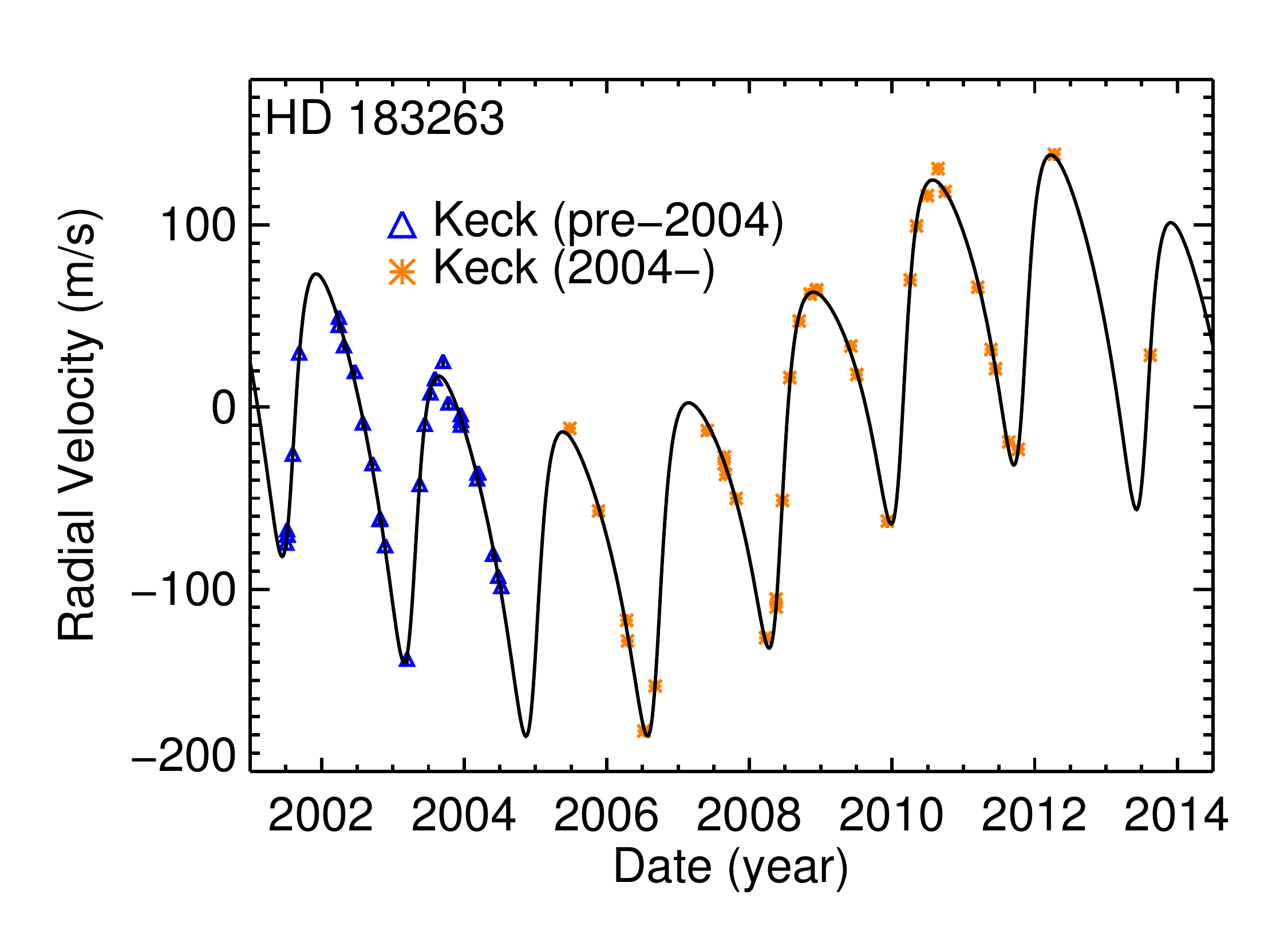}
    \label{183263sys}}
\quad
\subfigure[Residuals]{
    \includegraphics[scale=0.32]{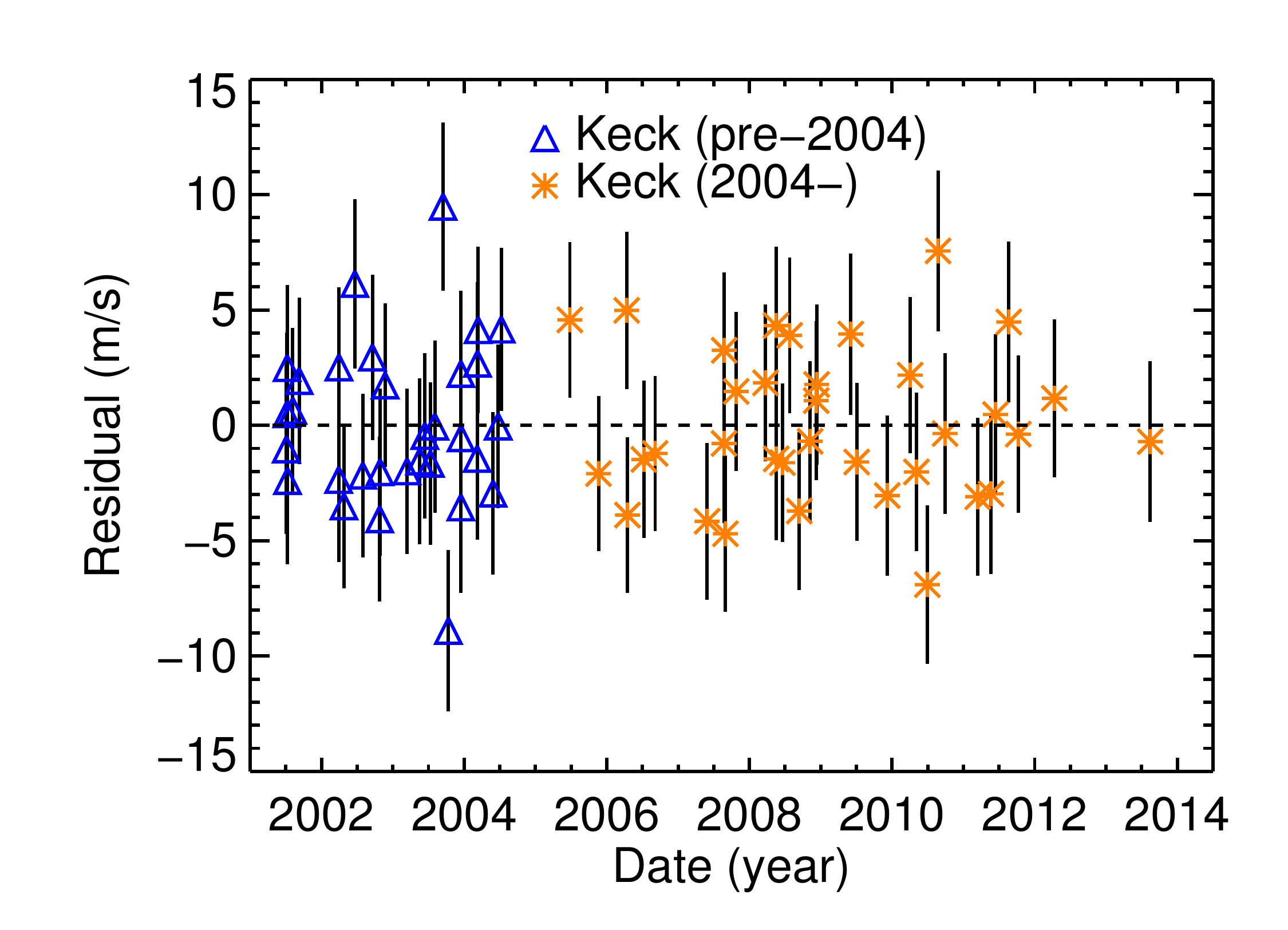}
    \label{183263res}}
\subfigure[HD 183263 \textit{b}]{
    \includegraphics[scale=0.32]{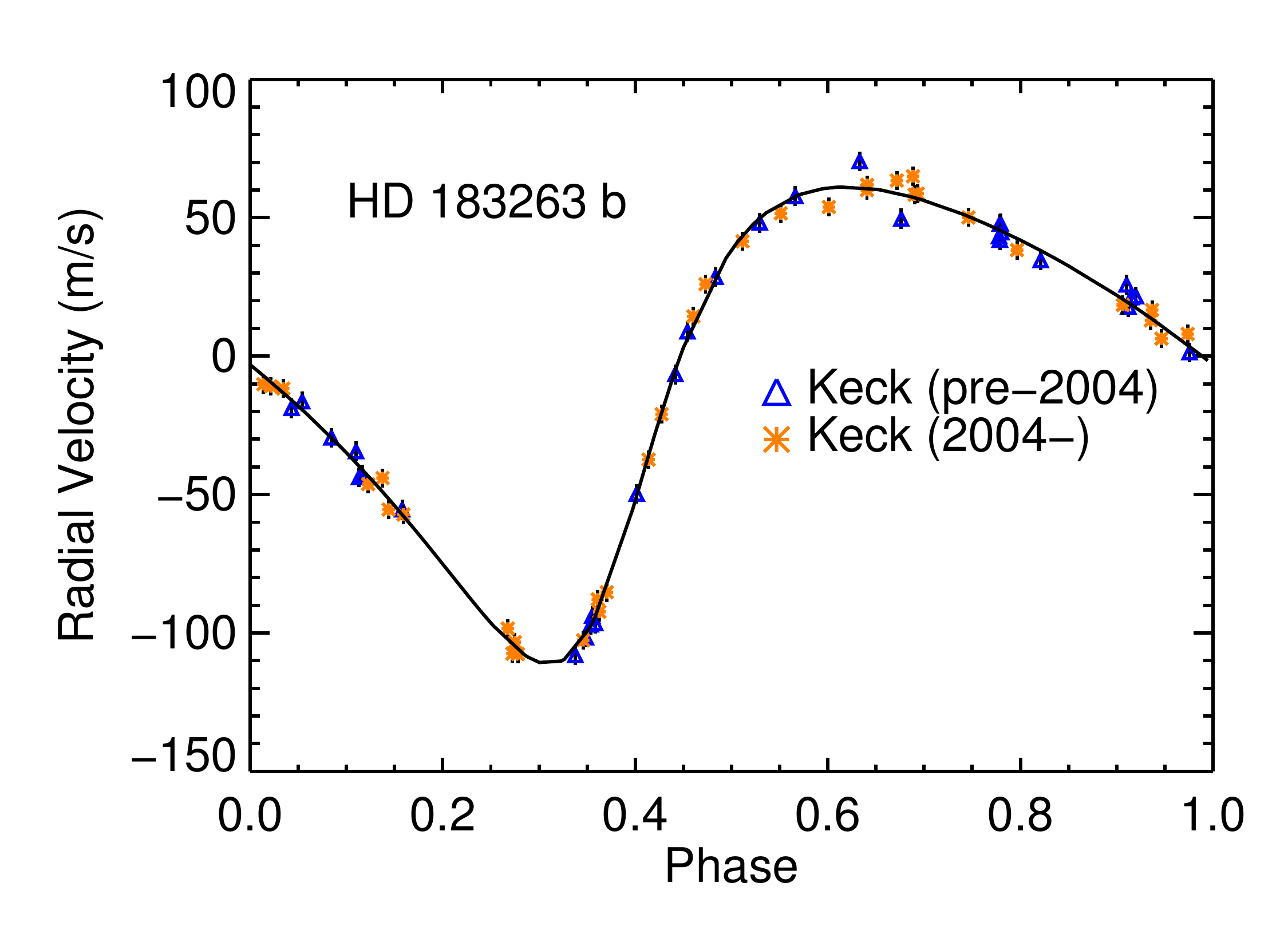}
    \label{183263b}}
\quad
\subfigure[HD 183263 \textit{c}]{
    \includegraphics[scale=0.32]{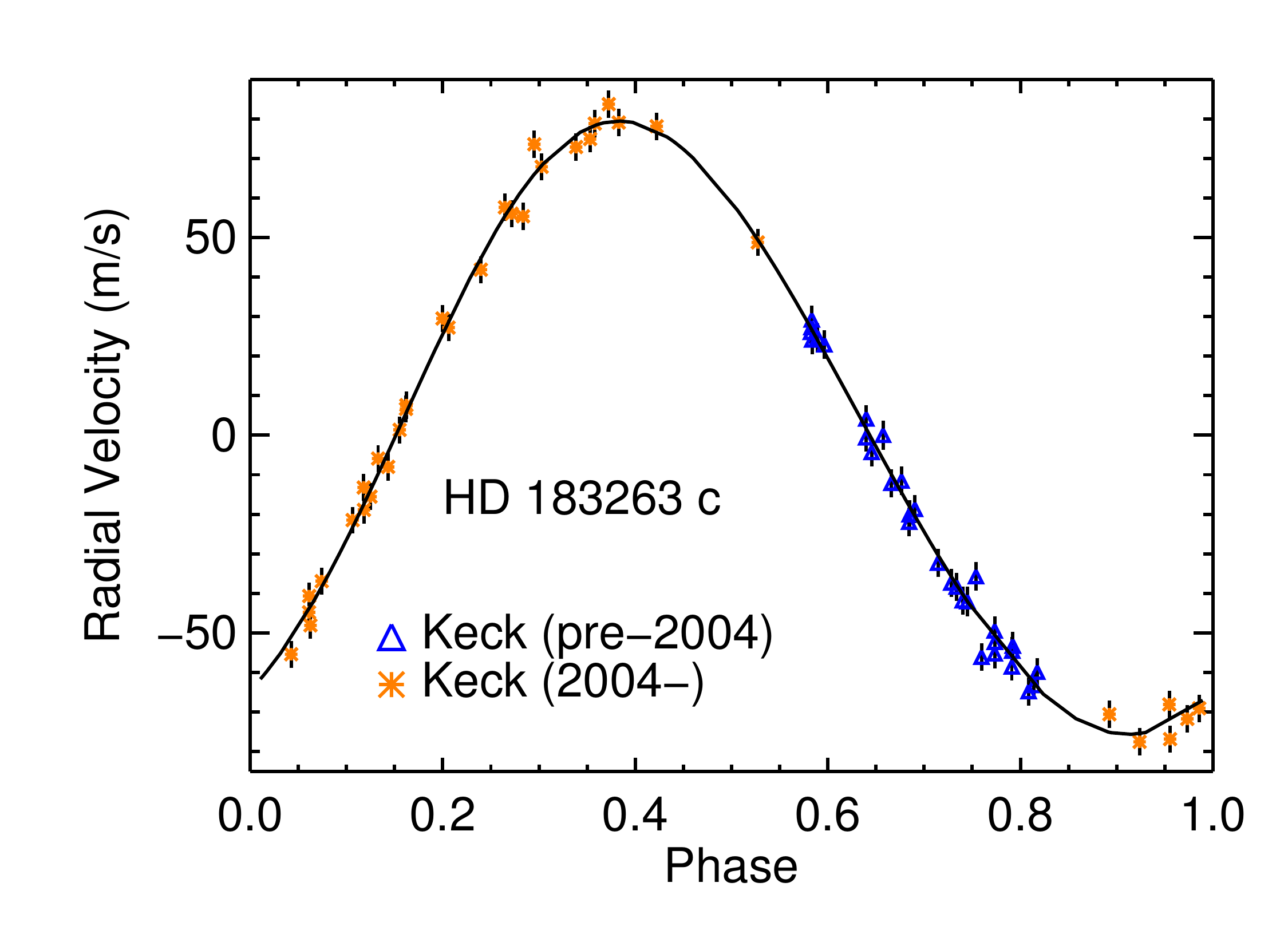}
    \label{183263c}}

\caption[Radial velocity and Keplerian fits for the HD 183263 system.]{Radial velocity and Keplerian fits for the HD 183263 system. Solid lines represent the best-fit Keplerian orbits. 

\ref{183263sys} Keck RVs overplotted by best-fit two-planet Keplerian model. \ref{183263res} Residuals of the RVs with the best-fit two-planet Keplerian model subtracted. \ref{183263b} and \ref{183263c}: the RV curves for HD 183263 \textit{b} and \textit{c}, respectively.} \label{183263plot}
\end{figure}

With 66 velocities from HIRES, we implemented a fit with an rms of 3.68 m s$^{-1}$ and an assumed jitter of 3.2 m s$^{-1}$. Figure \ref{183263plot} presents the RV curves for the system as well as the residuals. The orbit for HD 183263 \textit{c} appears to have finally closed, and it is significantly closer to circular ($e = 0.051 \pm 0.010$) and has a longer period than the solution from Wright et al. (2009), which found $e = 0.239 \pm 0.64$ and $P \sim 8.5$ yr. We find for HD 183263 \textit{c}, that $P = 4684 \pm 71$ days, or 9.1 yr; $M\sin i$ is $6.90 \pm 0.12 M_{\rm Jup}$. While our best fit orbital solution does not match well with the previous orbital solution, our solution resides comfortably within the stable portion in the $P_c$--$M_c\sin i_c$ space found by \citet[][see their Figure 3]{2009wright}.

\subsection{HD 187123}

\begin{figure}
\centering
\subfigure[HD 187123 system]{
    \includegraphics[scale=0.32]{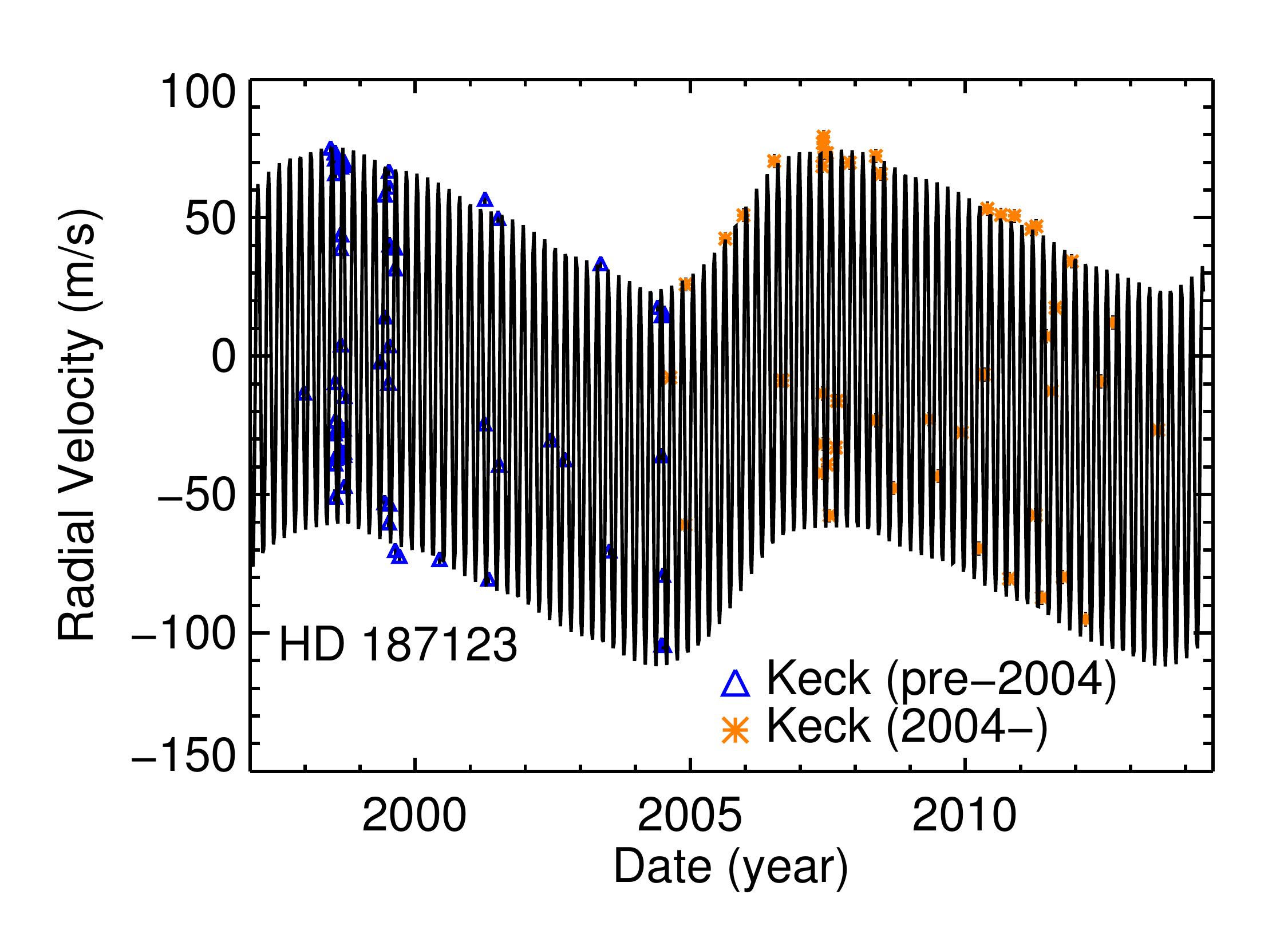}
    \label{187123sys}}
\quad
\subfigure[Residuals]{
    \includegraphics[scale=0.32]{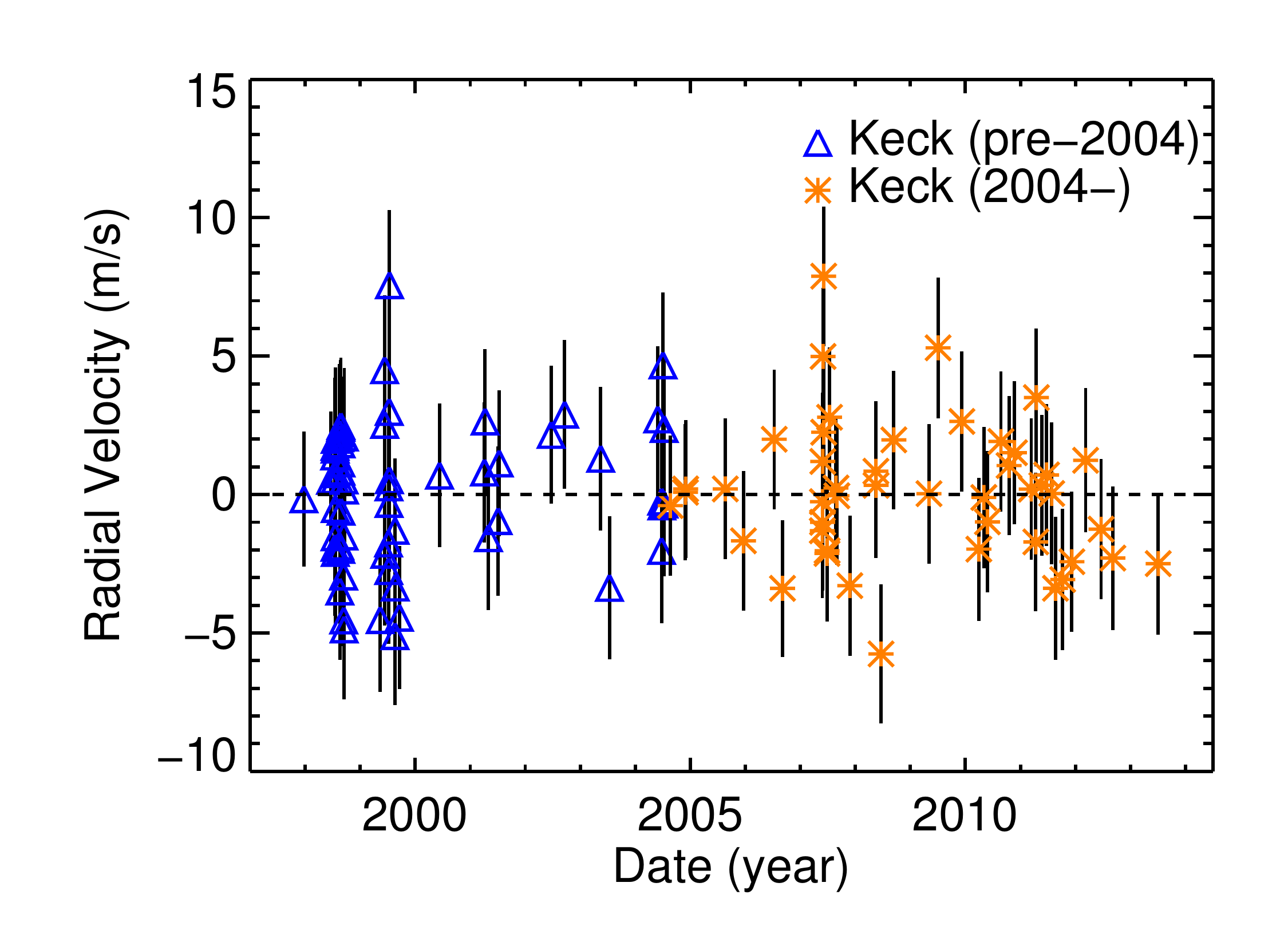}
    \label{187123res}}
\subfigure[HD 187123 \textit{b}]{
    \includegraphics[scale=0.32]{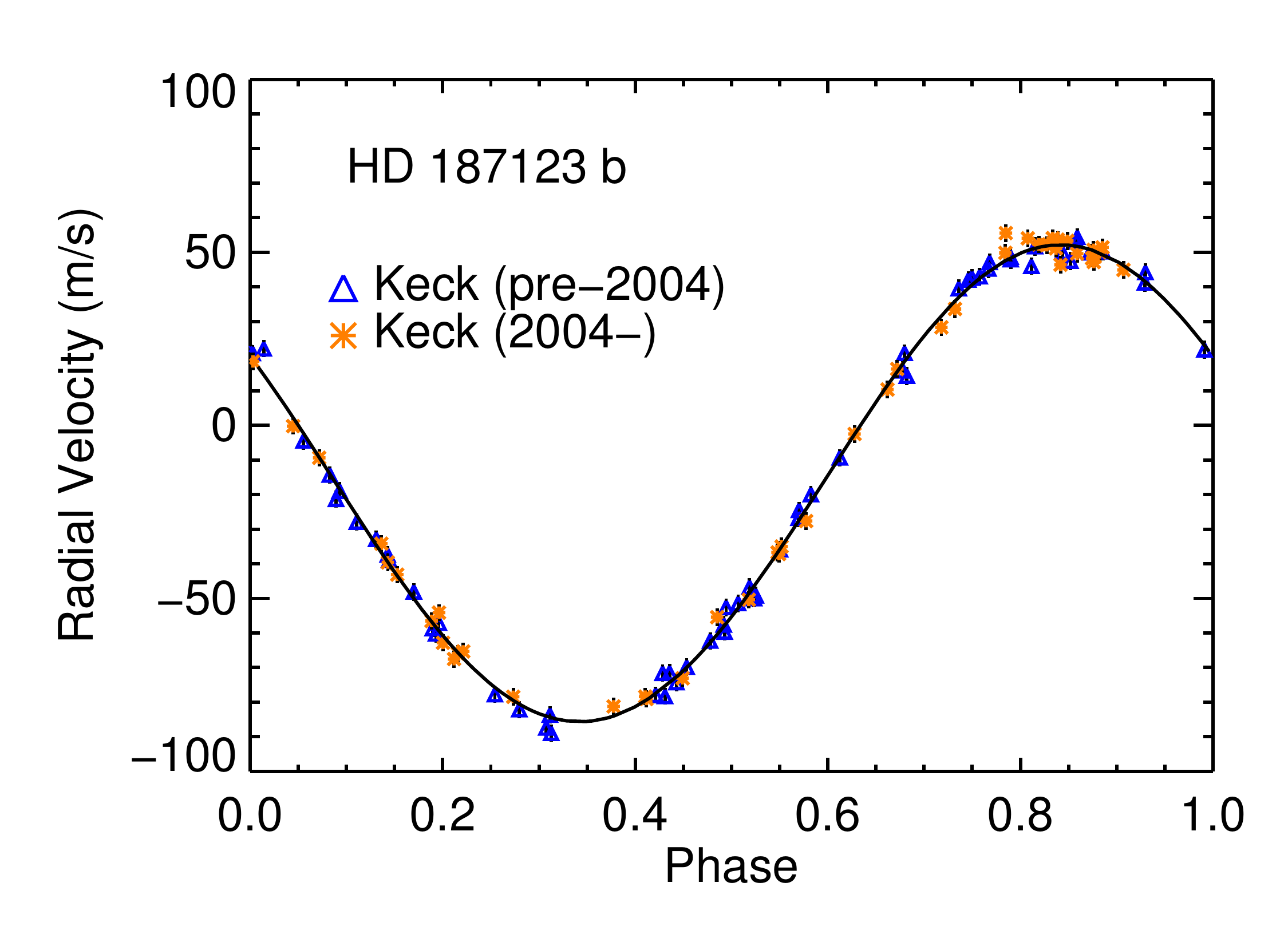}
    \label{187123b}}
\quad
\subfigure[HD 187123 \textit{c}]{
    \includegraphics[scale=0.32]{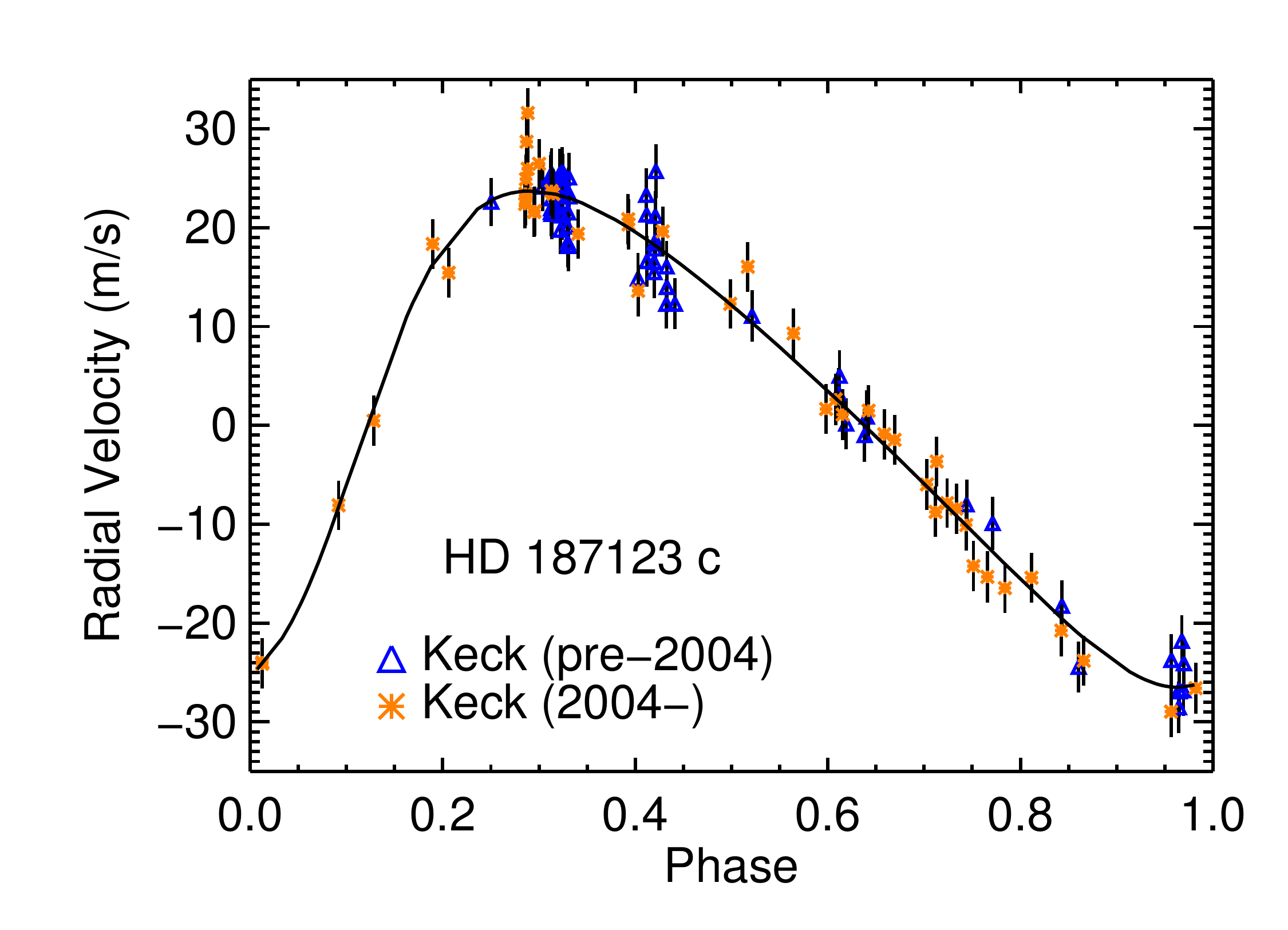}
    \label{187123c}}
\caption[Radial velocity and Keplerian fits for the HD 187123 system.]{Radial velocity and Keplerian fits for the HD 187123 system. Solid lines represent the best-fit Keplerian orbits. 

\ref{187123sys} Keck RVs overplotted by best-fit two-planet Keplerian model. \ref{187123res} Residuals of the RVs with the best-fit two-planet Keplerian model subtracted. \ref{187123b} and \ref{187123c}: the RV curves for HD 187123 \textit{b} and \textit{c}, respectively.}
\label{187123plot}
\end{figure}

\citet{butler1998} discovered HD 187123 \textit{b}, a $0.52 M_{\rm Jup}$ planet in a 3-day orbit. After many years of continued monitoring of this system, \citet{wright2007} announced a long-period outer companion with $P > 10$ yr and a minimum mass between 1.5 $M_{\rm Jup}$ and 10 $M_{\rm Jup}$. \citet{2009wright} presented a solution that constrained the mass and period of an outer companion to within 20\%, with $P = 10.4 \pm 1.2$ yr and $M \sin i = 2.0 \pm 0.3 M_{\rm Jup}$. Figure~\ref{187123plot} shows an updated fit with HIRES data. \citet{naef2004} provide ELODIE velocities; however, since they have significantly worse precision and do not add temporal coverage, we do not use them here. The 108 Keck observations still cover multiple orbits of the planets; assuming a jitter of 2.23 m s$^{-1}$, we find an rms of 2.66 m s$^{-1}$. From our fit, the period of HD 187123 \textit{c} is $9.1 \pm 0.13$ yr and the minimum mass is $1.818 \pm 0.035 M_{\rm Jup}$. HD 187123 \textit{c} appears to be a Jupiter analog, although its orbit is somewhat eccentric at $e = 0.280 \pm 0.022$.

\subsection{HD 217107}

\citet{fischer1999} presented HD 217107 \textit{b} as a 1.27 $M_{\rm Jup}$ planet in a 7.12-day period. A few years later, \citet{fischer2001} identified a linear trend in the residuals, which was likely caused by an outer companion. \citet{vogt2005} reported the first orbit for HD 217107 \textit{c}, modestly constrained at $P = 8.6$ yr and $M\sin i = 2.1 M_{\rm Jup}$. \citet{2009wright} constrained the orbit and mass of HD 217107 \textit{c} to almost within 10\%, with $P\sim11.7$ yr and the minimum mass $\sim 2.6 M_{\rm Jup}$.

\begin{figure}
\centering
\subfigure[HD 217107 system]{
    \includegraphics[scale=0.32]{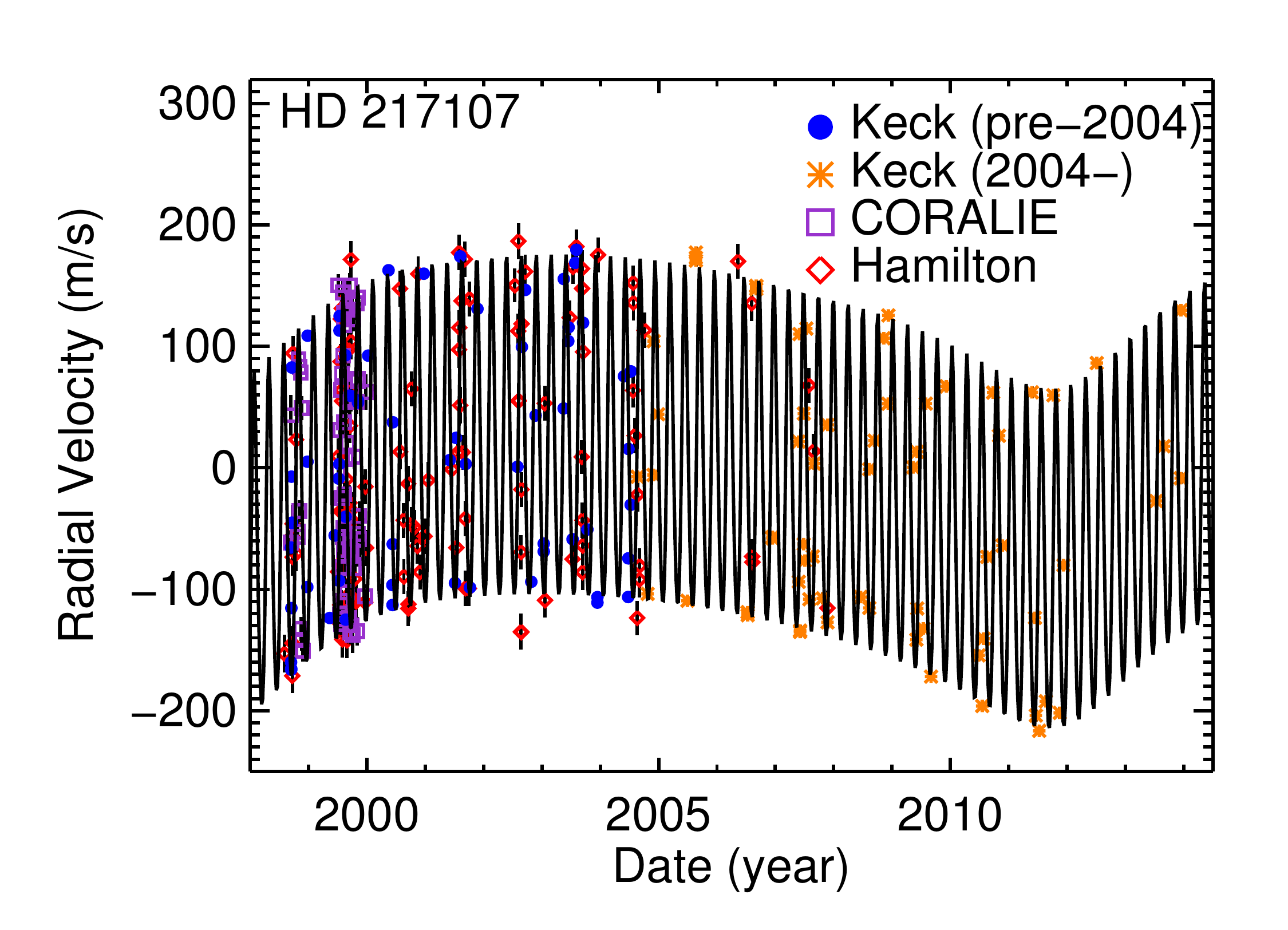}
    \label{217107sys}}
\quad
\subfigure[Residuals]{
    \includegraphics[scale=0.32]{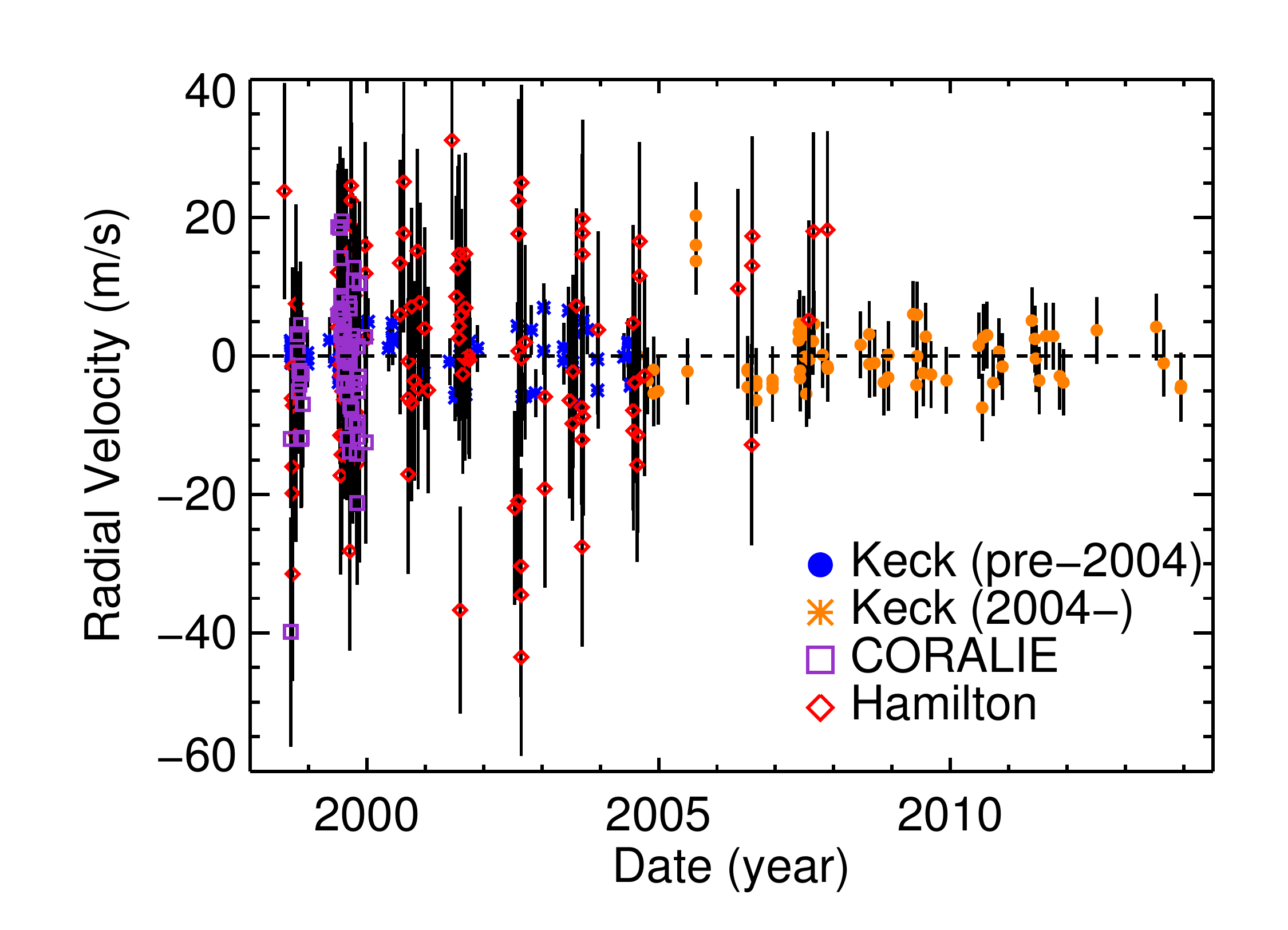}
    \label{217107res}}
\subfigure[HD 217107 \textit{b}]{
    \includegraphics[scale=0.32]{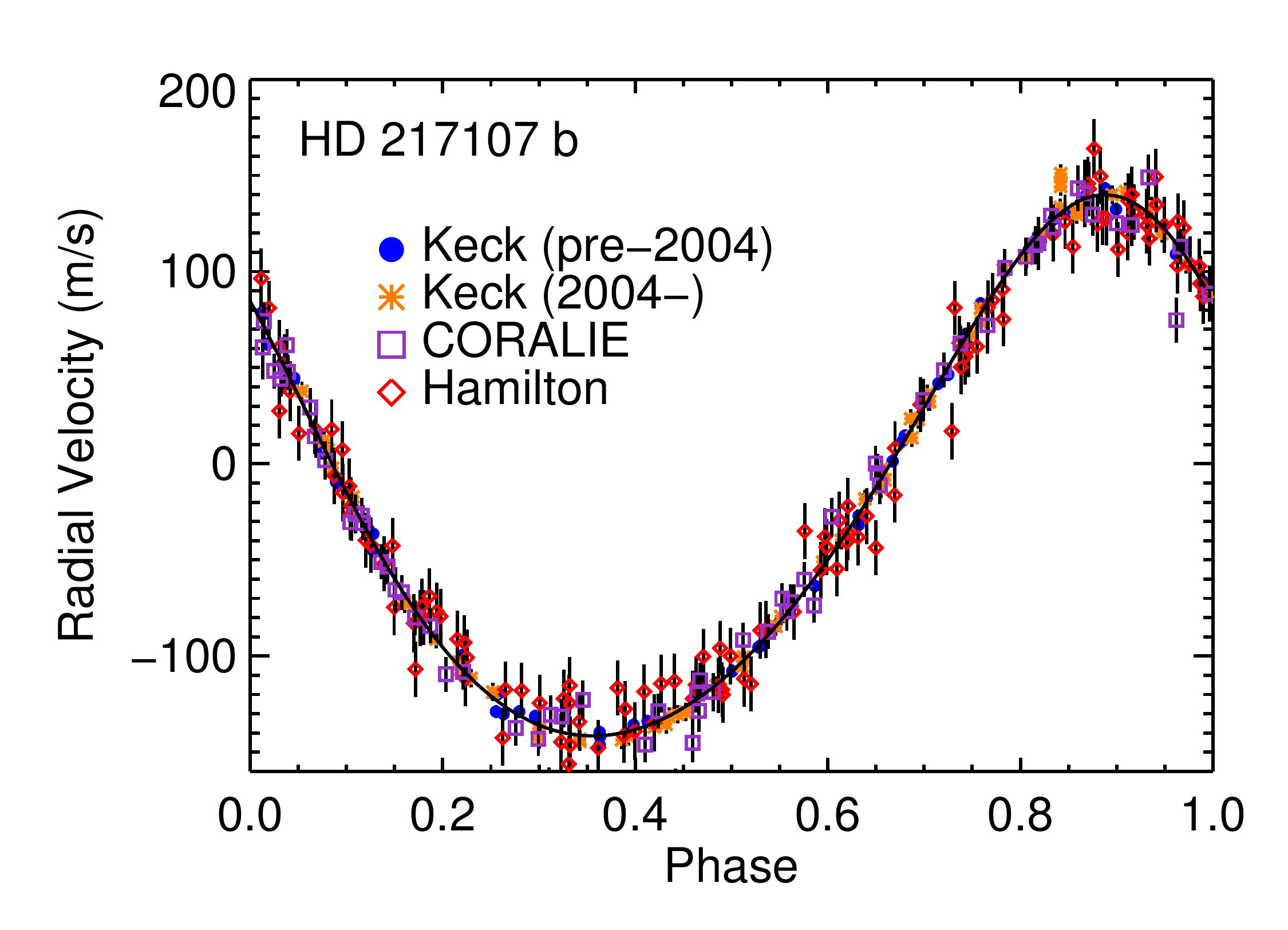}
    \label{217107b}}
\quad
\subfigure[HD 217107 \textit{c}]{
    \includegraphics[scale=0.32]{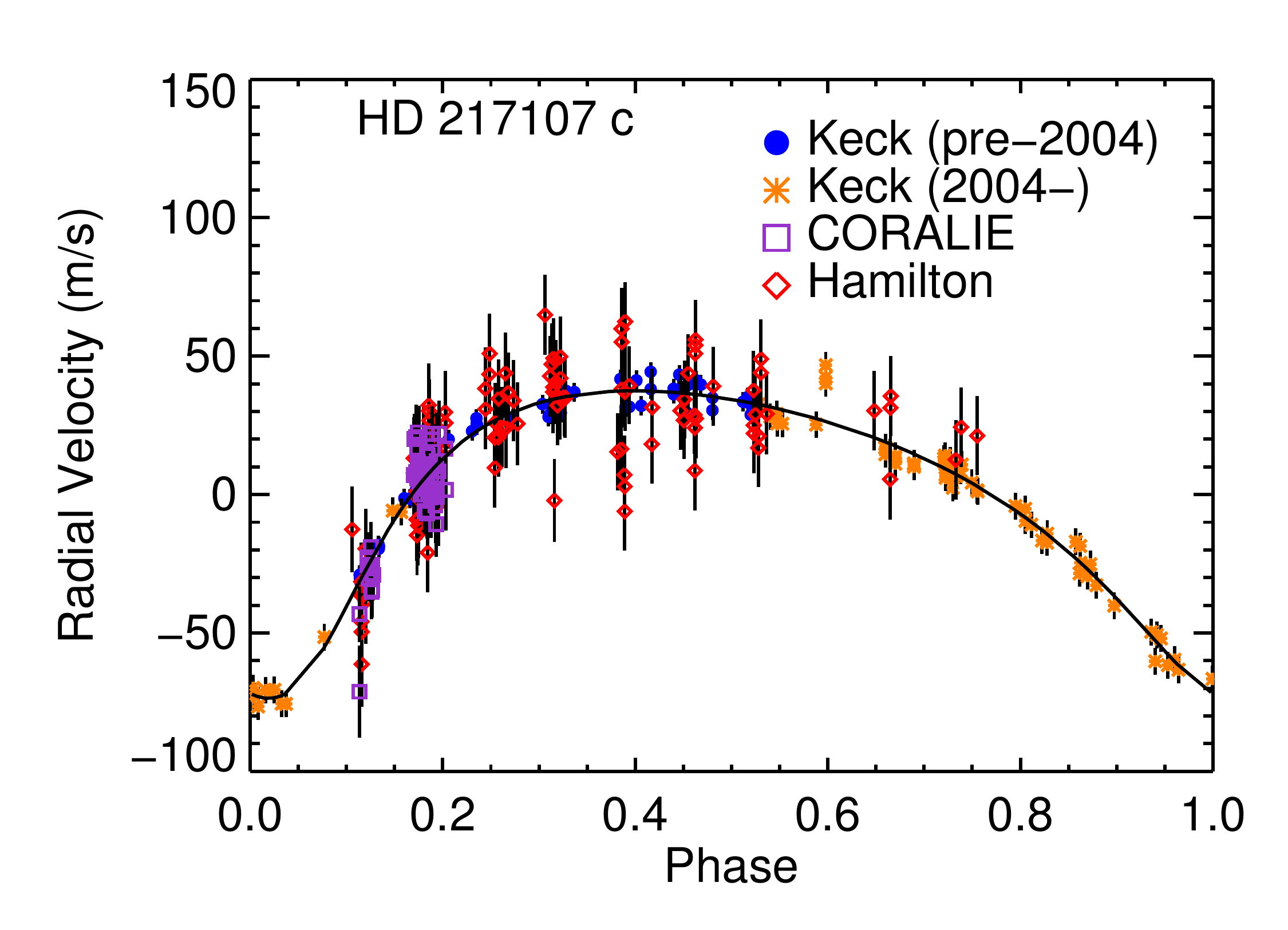}
    \label{217107c}}
\caption[Radial velocity and Keplerian fits for the HD 217107 system.]{Radial velocity and Keplerian fits for the HD 217107 system. Solid lines represent the best-fit Keplerian orbits.

\ref{217107sys} Keck RVs overplotted by best-fit two-planet Keplerian model. \ref{217107res} Residuals of the RVs with the best-fit two-planet Keplerian model subtracted. \ref{217107b} and \ref{217107c}: the RV curves for HD 217107 \textit{b} and \textit{c}, respectively.}
\label{217107plot}
\end{figure}

As with the case of HD 74156, we also have data taken by different teams from several instruments, we added jitter instrument-by-instrument. In our fit, we use velocities from Keck (128 observations), Lick \citep[121]{2009wright}, and CORALIE \citep[63][]{Naef2001} to find a fit an rms of 10.29 m s$^{-1}$ (see Figure~\ref{217107plot}).  

Because the outer planet has only barely (apparently) completed an orbit, its orbital parameters may be especially uncertain (and are particularly sensitive to the assumption that there is not a third, longer-period planet contributing significantly to the velocities).  To explore the robustness of our derived orbital period of the $c$ component as a function of its minimum mass, we have constructed a $\chi^2$ map in $P$--$M\sin{i}$ space (a variety of what \citet{Knutson14} call ``Wright diagrams''; see \citet{Patel2007}, \citet{2009wright} and similar approaches taken in, e.g., \citet{Dumusque2011,boisse2012}).  In this map all orbital parameters have been optimized (i.e., they are at their maximum likelihood in a \kai\ minimum sense) for each pair of $P_c$ and $M_c \sin{i_c}$ in the map (except for the offsets among the four instruments, which are held constant at their overall best-fit values).

\begin{figure}
\begin{center}
\plotone{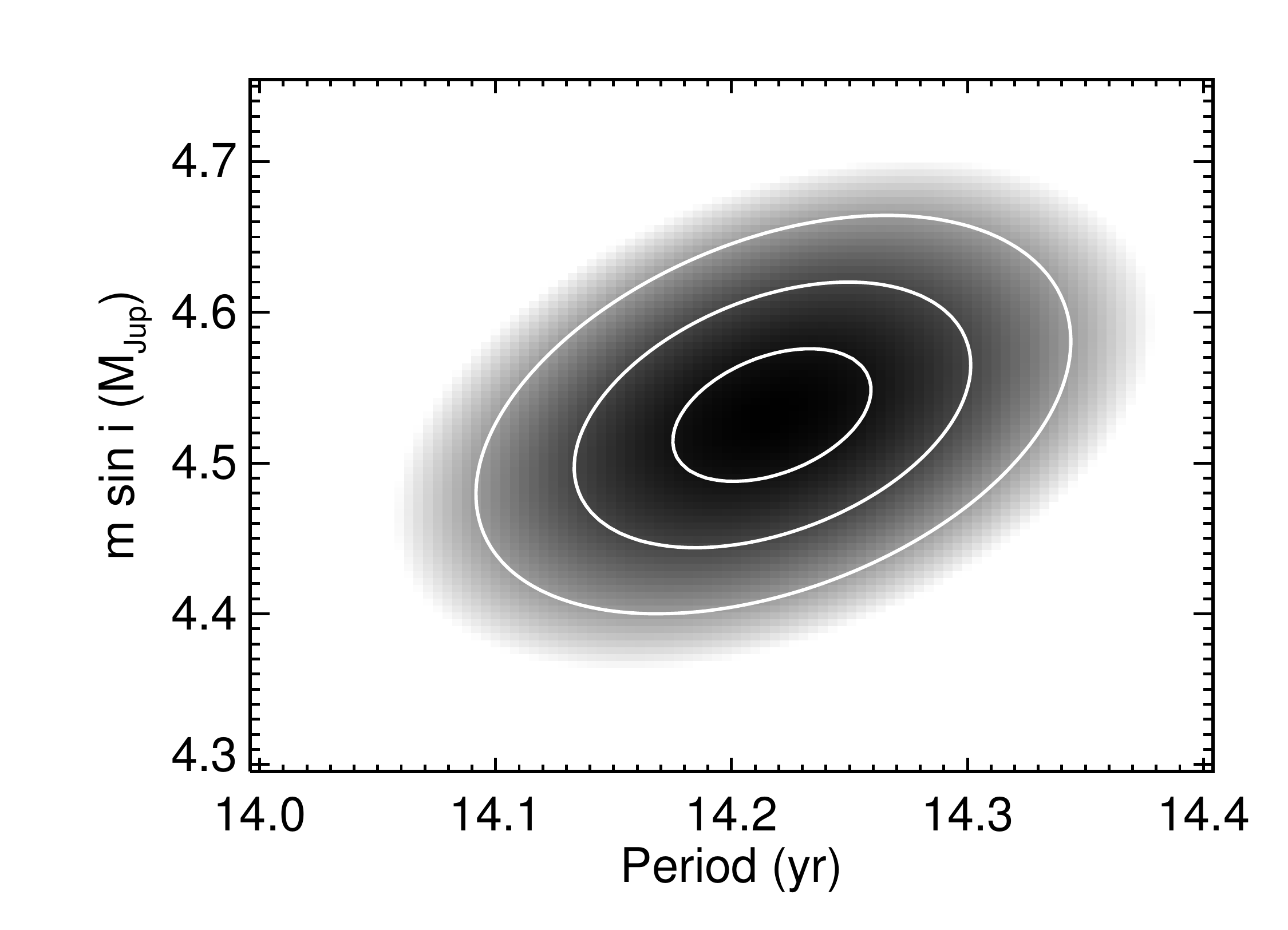}
\end{center}
\caption[Best-fit $100 \times 100$ $\chi^2$ map for HD 217107 c]{Best-fit $100 \times 100$ $\chi^2$ map for fixed values of \pc and \mc for HD 217107 \textit{c}. This confirms that the period and mass are well-constrained. We have illustrated the contours of the $1\sigma$, $2\sigma$, and $3\sigma$ (defined by $\chi^2=\chi^2_{\rm min} +\{2.30,6.17,11.8\}$) confidence levels, based on for the number of degrees of freedom in the problem \citep{NRC}.  The center and $1\sigma$ limits in both parameters are consistent with the bootstrapping uncertainties for these parameters.}
\label{217107chi2}
\end{figure}

Figure \ref{217107chi2} shows the \kai\ contour map, revealing that the orbital period and minimum mass for HD 217107 \textit{c} are well constrained with $P = 14.215_{-0.04}^{+0.045}$ yr and $M\sin i = 4.51_{-0.02}^{+0.07} M_{\rm Jup}$.  These uncertainties are roughly consistent with the uncertainties determined via bootstrapping, which yields $P=14.215\pm0.06$ yr and $M\sin i = 4.51\pm 0.07 M_{\rm Jup}$.  This validates our choice of stellar jitter for this star, since the contours in the \kai\ maps are sensitive to the choice of jitter, while the bootstrapping uncertainties are almost completely independent of it.  We report the bootstrapping uncertainties in Table~\ref{orb}.

To test the importance of our assumption that there are only two planets contributing detectable accelerations to the star, we repeated our bootstrapping analysis with a model that includes an additional, linear trend to the data.  Though there is no statistical need to include such a trend in our model, giving our model the freedom to include one could, in principle, affect the best-fit parameters for the outer planet.  Indeed, though the parameters of the $b$ component do not change significantly in this model (as expected given its high frequency) we find a slightly different best-fit with such a model, with $P_c, T_{{\rm p},c}$, and $K_c$ all changing by 2--4 $\sigma$, resulting in a minimum mass for the outer companion of $4.37 M_{\rm Jup}$. The uncertainties on the parameters of the $c$ component in the with-trend model are larger by a factor of 2--4, comfortably including most of the parameter estimates from the no-trend model.   We conclude that our choice not to include a linear trend does not have a large effect on our conclusions or parameter estimations.

\subsection{GJ 849}

\subsubsection{Orbital Fit}

Unlike the other stars in this work, GJ 849 is an M3.5 dwarf.  Various studies of this star's composition have all found similar, super-solar abundances:  \citet{rojas2012} find [Fe/H] = 0.31$\pm$0.17 (from $K$-band features); \citet{onehag2012} find 0.35$\pm$0.10 (using $J$-band); and \citet{terrien2012} found 0.31$\pm$0.12 (using $K$-band). 

GJ 849 hosts the first planet discovered orbiting an M-dwarf with a semi-major axis greater than 0.21 AU. \citet{butler2006} announced GJ 849 \textit{b}, with $P= 5.16$ yr and minimum mass $0.82 M_{\rm Jup}$. At the time, there was evidence of a linear trend of $-4.75$ m s$^{-1}$ yr$^{-1}$, indicative of a second companion. \citet{bonfils2013} also fitted the system with one planet and a linear trend of $-4.0$ m s$^{-1}$ yr$^{-1}$, adding their HARPS data to the published HIRES velocities. 

Stellar magnetic activity had to be ruled out as the source of the trend. \citet{gomes2012} monitored several M-dwarfs from the HARPS program for long-term magnetic activity. For GJ 849, they saw mild correlation in our velocities with the \ion{Na}{1} index data. However, the amplitude was not large enough. \citet{montet2014} provided the first orbital parameters for on GJ 849 \textit{c}, finding $M\sin i = 0.70 \pm 0.31 M_{\rm Jup}$, and $P = 19.3_{-5.9}^{+17.1}$ yr, and found no correlation between stellar magnetic activity and the long-period signal of this outer companion.

Our fit, using 35 velocities from HARPS \citep{bonfils2013} and 82 velocities from HIRES spanning from 1997 through early 2014, has further constrained the orbital parameters of the GJ 849 system. We incorporate a jitter of 3 m s$^{-1}$, and our fit has an rms of 3.72 m s$^{-1}$. 

GJ 849 \textit{b} is a $0.911 M_{\rm Jup}$ planet in a 5.27 yr period with an orbital eccentricity of 0.038. GJ 849 \textit{c} is a $0.944 \pm 0.07 M_{\rm Jup}$ planet in a $15.1 \pm 0.66$ yr period with an orbital eccentricity of $0.087 \pm 0.06$. 

GJ 849 \textit{c} has the longest robustly measured orbital semimajor axis of any planet orbiting an M dwarf discovered to date.  Indeed, it has one of the longest well-measured periods of exoplanets orbiting any kind of star.  Exoplanets with similar period and period uncertainties in the literature include 55 Cnc $d$ \citep{marcy2002,endl2012}; HD 166724 $b$ and HD 219077 $b$ \citep{Marmier2013}; and HD 13931$b$ \citep{CPS1} --- but these all orbit stars with $M>0.8 M_{\odot}$ and the two from \citeauthor{Marmier2013} show significant eccentricity.  The exoplanet with the longest robustly measured orbital period is $\beta$ Pictoris \citep[$P=20.5^{+2.9}_{-1.4}$ yr][]{Macintosh2014}.

\begin{figure}[H]
\centering
\subfigure[GJ 849 system]{
    \includegraphics[scale=0.32]{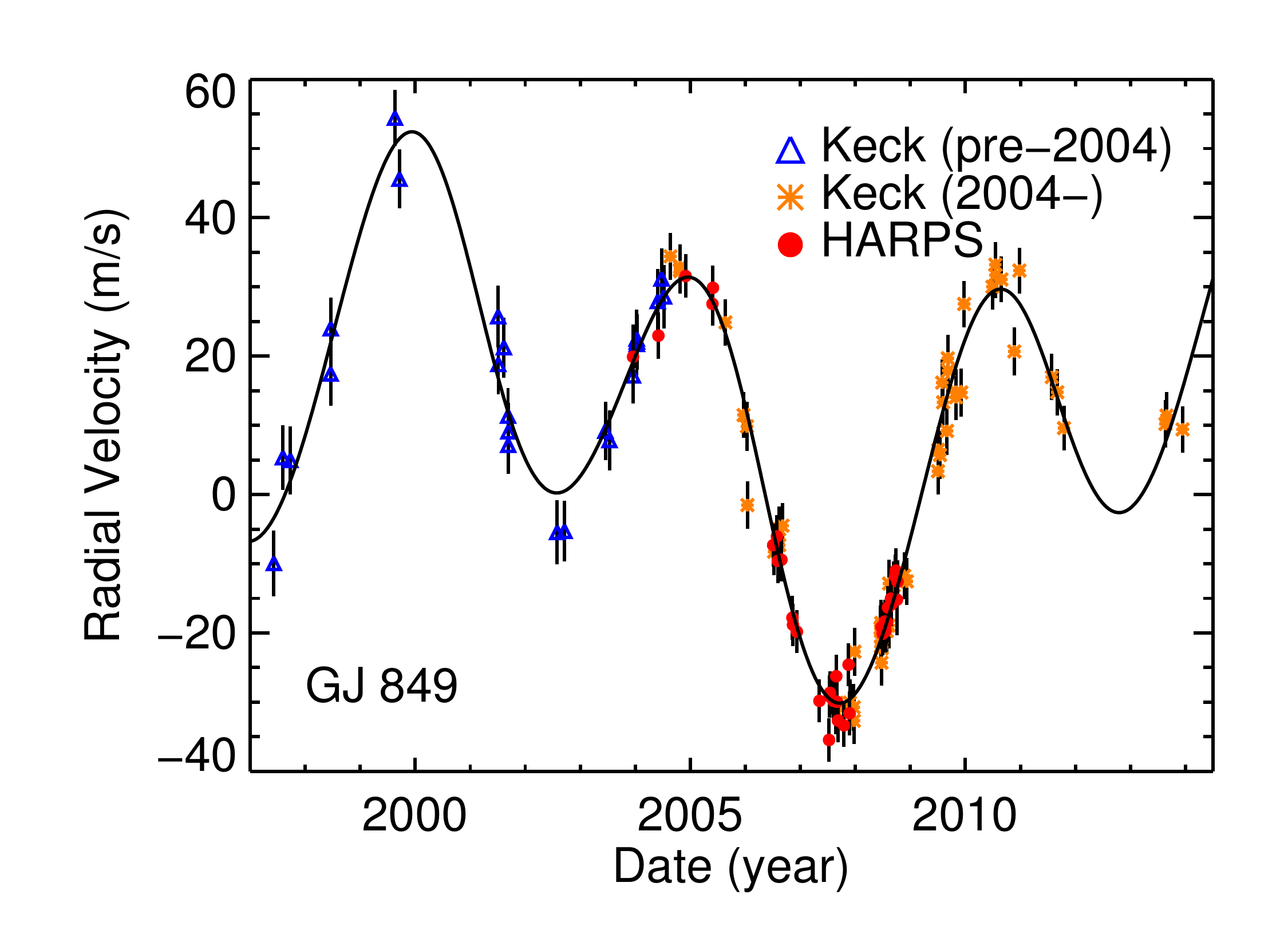}
    \label{849sys}}
\quad
\subfigure[Residuals]{
    \includegraphics[scale=0.32]{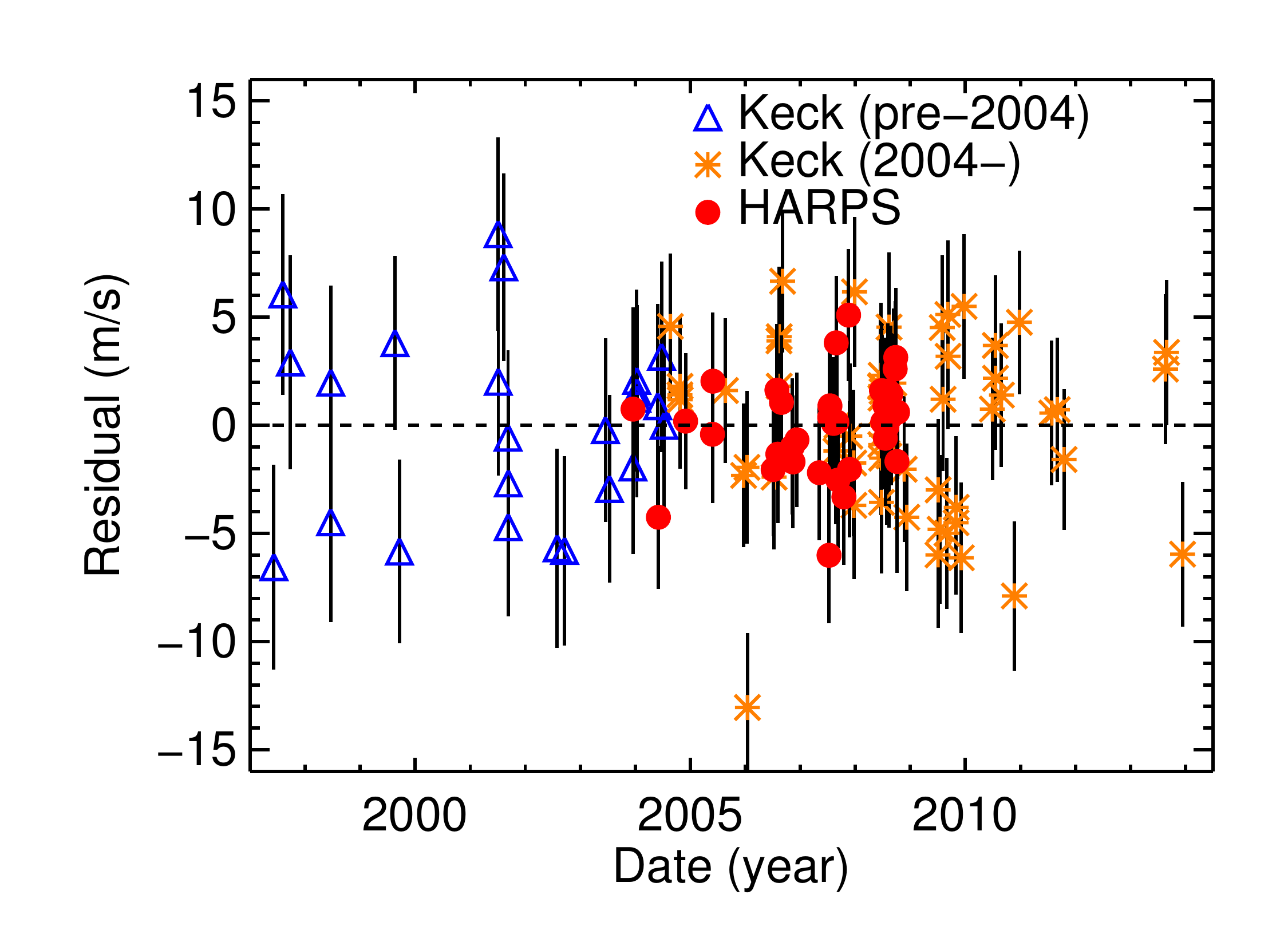}
    \label{849res}}
\subfigure[GJ 849 \textit{b}]{
    \includegraphics[scale=0.32]{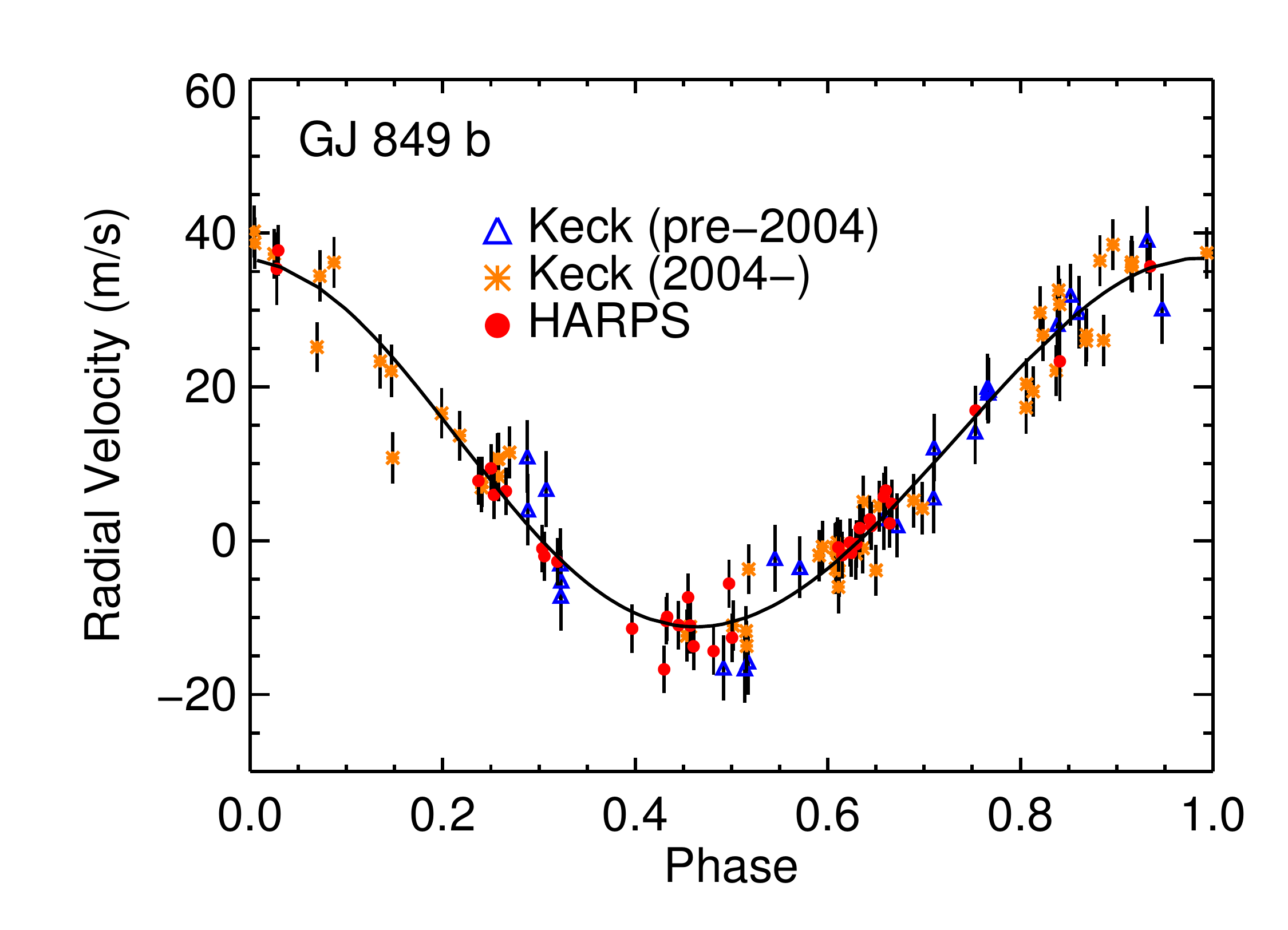}
    \label{849b}}
\quad
\subfigure[GJ 849 \textit{c}]{
    \includegraphics[scale=0.32]{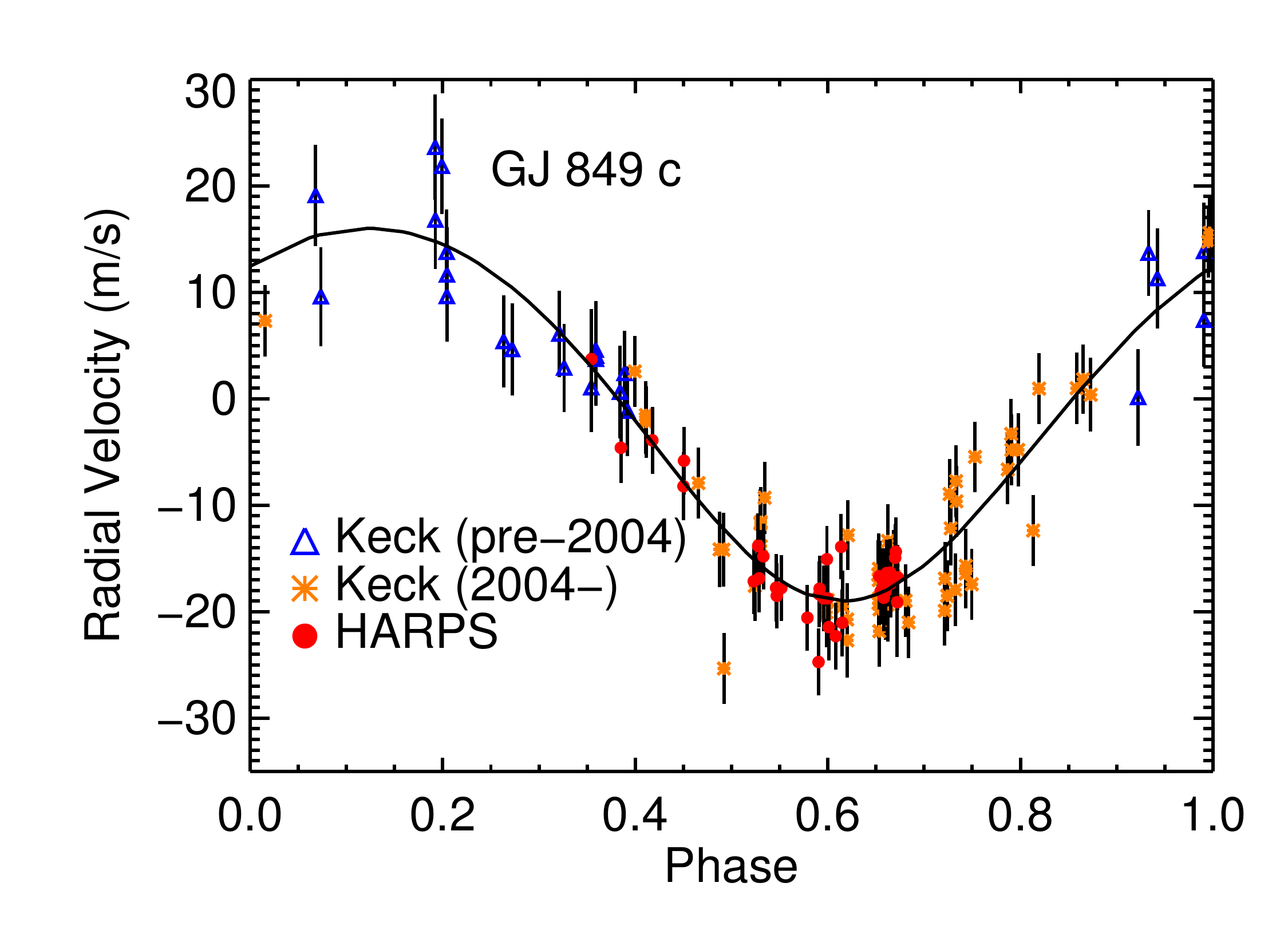}
    \label{849c}}
\caption[Radial velocity and Keplerian fits for the GJ 849 system.]{Radial velocity and Keplerian fits for the GJ 849 system. Solid lines represent the best-fit Keplerian orbits. 

\ref{849sys} Keck RVs overplotted by best-fit two-planet Keplerian model. \ref{849res} Residuals of the RVs with the best-fit two-planet Keplerian model subtracted. \ref{849b} and \ref{849c}: the RV curves for GJ 849 \textit{b} and \textit{c}, respectively.}
\label{849plot}
\end{figure}
 
We estimated the model parameters for GJ 849 in two additional ways to check for consistency and robustness. It is unclear whether the bootstrap resampling procedure provides an accurate estimate of GJ 849 \textit{c}'s orbital parameters. In particular, the poor phase coverage before 2001 (see Figure \ref{849plot}) results in several clear outlier models in the joint parameter distributions. 

\begin{figure}
  \plotone{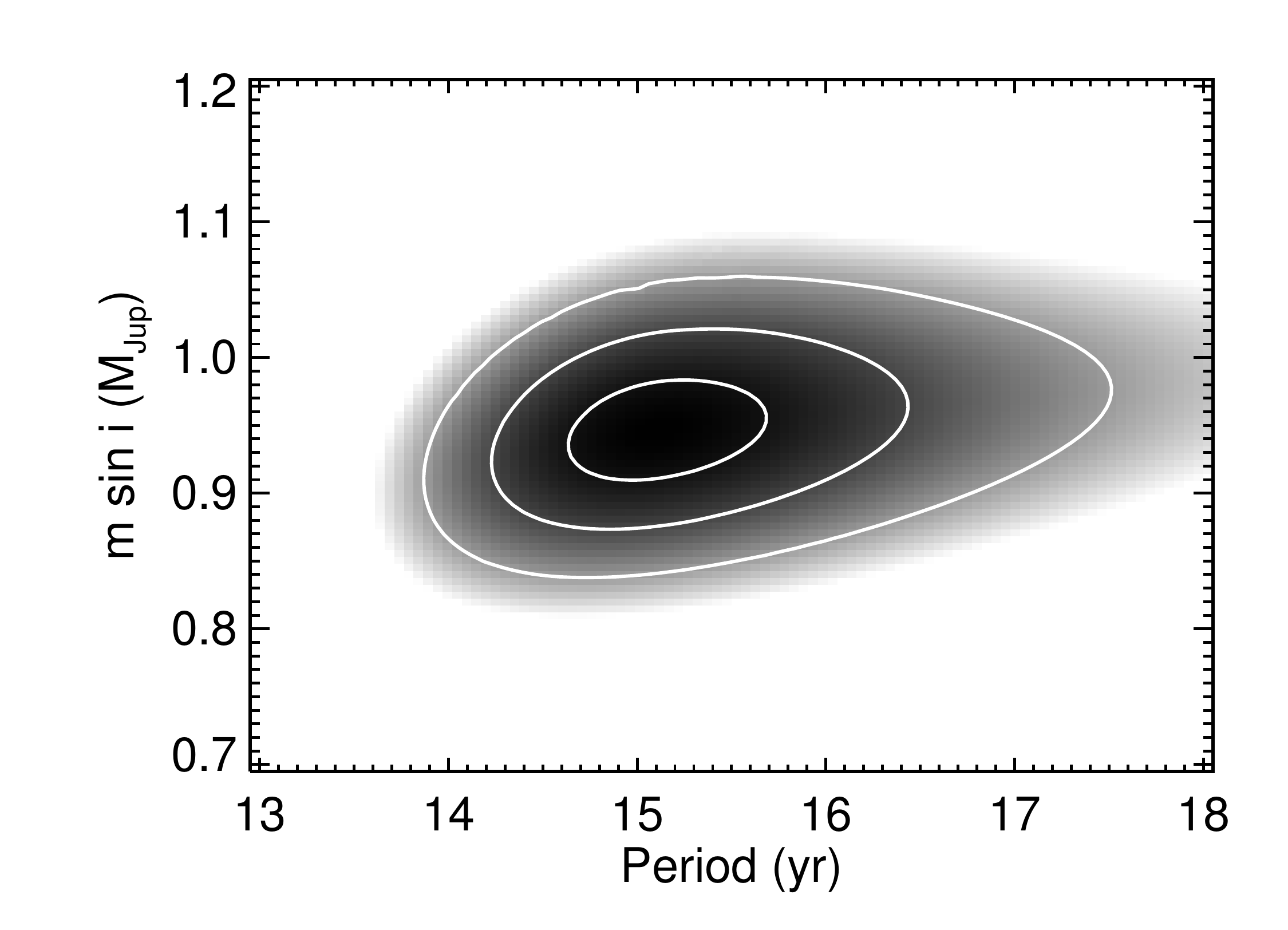}
\caption{Contours at $1\sigma$, $2\sigma$, and $3\sigma$ (defined by $\chi^2=\chi^2_{\rm min} +\{2.30,6.17,11.8\}$) confidence levels, as appropriate for the number of degrees of freedom in the problem \citep{NRC} for the orbital parameters of GJ 849$c$.  The period and mass of the $c$ component appear to be well constrained it better than 5\%.\label{fig:GJ849_contour} }
\end{figure}

In our first check for accuracy in the parameters and uncertainties, as with HD 217107 \textit{c}, we constructed a $P$--$M \sin{i}$ \kai\ map to confirm that the orbital period of GJ 849 \textit{c} is well constrained (assuming no additional planets and a stellar jitter of 3 m s$^{-1}$), despite having just completed an orbit, and find that the 68\% confidence interval contours corresponds to uncertainties in $P$ of less than 5\%.  As Figure \ref{fig:GJ849_contour} shows, the \kai\ map uncertainties in minimum mass are 0.07 $M_{\rm Jup}$, exactly consistent with our bootstrapping errors; the uncertainties in period are $\sim 1.1$ yr, which is larger than the bootstrapping errors of 0.66 yr, probably because the \kai\ contours are asymmetric.  

As a second check, we turn to a Bayesian approach for performing parameter estimation via Markov chain Monte Carlo. We adopt the usual broad priors for Keplerian orbital parameters and likelihood assuming uncorrelated, Gaussian measurement errors with dispersion based on the quadrature sum of the reported measurement uncertainties and an unknown jitter term \citep{Ford06}.  Given the potential for mutual planetary interactions, we apply RUN DMC\footnote{We used the Keplerian parameter priors given in \citet{Nelson14a}, and the algorithmic parameters $n_{\rm chains}$=300, $n_{\rm gen}$=100,000, $\sigma_\gamma$=0.01, and MassScaleFactor=1.0.}, a well tested code that combines n-body integration with differential evolution Markov chain Monte Carlo \citep{Nelson14a}. Although the GJ 849 planets are well approximated by Keplerian orbits, the differential evolution proposal in RUN DMC is much more efficient than a traditional random walk MCMC for dealing with correlated parameters, which are often present in the parameters for long-period companions, and so by using RUN DMC we do not have to fine tune a proposal distribution.  We find that the marginal posterior probability distribution for $P_c$ has 68\% of its mass within $0.74$ yr of the median period of 15.1 yr, only slightly larger than the uncertainty estimated from the bootstrap.  

The similarity of the parameter uncertainties from all three methods verifies that the orbit of GJ 849 \textit{c} is well constrained and validates the BOOTTRAN and \kai\ map approaches (in this case) and our choice of jitter.  We use the more conservative \kai\ contours do determine parameter uncertainties in Table~\ref{orb}.  

\subsubsection{Stability}

Because this system is not ``highly hierarchical'' \citep{wright2010} in mass or orbital period, we have performed $n$-body simulations to establish the dynamical stability of our orbital solutions.   The 1000  \BOOTTRAN realizations of the GJ 849 RV data are used to determine parameter uncertainties is associated with a complete set of Keplerian orbital parameters for the two planets ($P$, $e$, $\omega$, $K$, and $T_{\rm p}$ for each planet, plus an overall RV offset $\gamma$ and two offsets among the three RV data sets).  All of these realizations returned reasonable fits, indicating that the fitting procedure did not fail in any case.  

We performed long-term dynamical integrations for all 1000 fits to these realizations of the data  using the \MERCURY symplectic integrator \citep{Chambers99}.  Each simulation runs for $10^7$ orbits of the inner-most planet ($\sim2\times10^{10}$ days). This integration timescale is short relative to the lifetime of the star but sufficiently long enough to show a significant fraction of our models undergo an instability, described below. 

An instability occurs if at any point during the integration either planet crosses the other's Hill sphere or either of the planets' semi-major axes change by more than 50\% of their initial value. 

\begin{figure}
\plotone{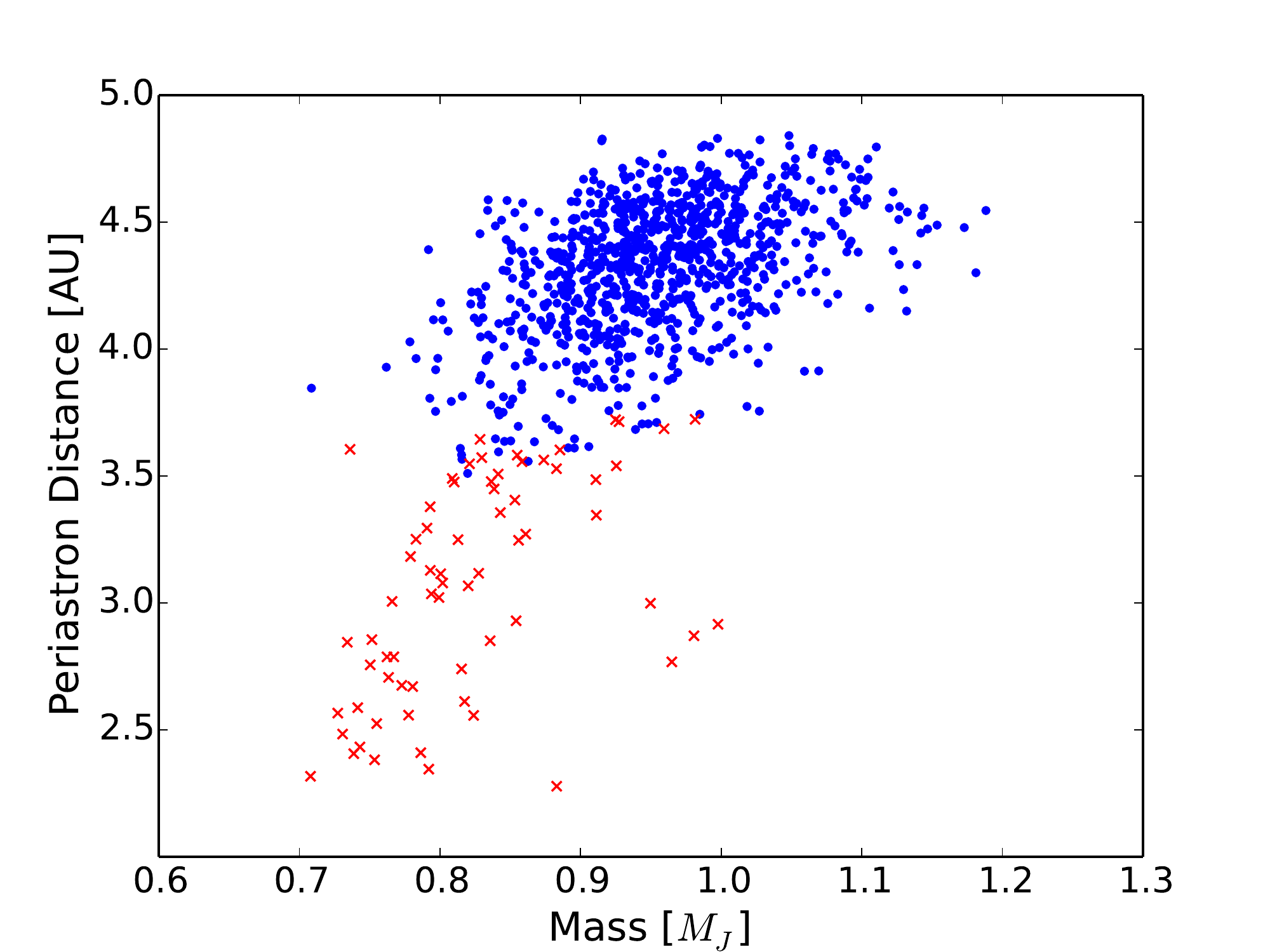}
\caption{Best fit minimum mass and periastron distance for all of the 1000 \BOOTTRAN\ realizations of the GJ 849 system (which were generated through a bootstrapping procedure performed on the radial velocity data).  Red x's represent fits which proved to be unstable in fewer than $10^7$ orbits.  There is a clear boundary at periastron distance $\sim 3.6$ (AU) between stable and unstable configurations.  We have not used the unstable realizations in determining parameter uncertainties for the GJ 849 system. \label{fig:GJ849stability}}
\end{figure}

None of our models resulted in a collision over the course of the integration. However, we find 67 models undergo the second listed mode of instability ($\mid[a_{\rm final}-a_{\rm initial}]/a_{\rm initial}\mid > 0.5$) when the periastron passage of GJ 849 $c$ is less than 3.5 AU (Figure \ref{fig:GJ849stability}).  The instability times are logarithmically uniform from $\sim$10 to $\sim 10^7$ yr.

We removed the unstable \BOOTTRAN realizations from our calculations of the uncertainties in the orbital parameters we report in Table~\ref{orb}.

\section{A $2 M_{\rm Jup}$ Planet around HD 145934} \label{giant}

We here announce a new long-period planet orbiting the giant star HD 145934, a 1.748$\pm$0.105 $M_{\odot}$ star \citep{takeda2007}.  This star was not known to be a giant when the California Planet Survey began monitoring it in 1997 at Keck Observatory. Since then, its log(g) value and mass from \citet{takeda2007} indicate that it is a giant. Visual inspection of the gravity sensitive sodium and magnesium lines confirm this diagnosis. Radial velocities of HD 145934 show a clear sinusoidal modulation of planetary amplitude upon a large linear trend, indicative of a stellar binary companion.

\begin{figure}[H]
\centering
\subfigure[HD 145934]{
    \includegraphics[scale=0.32]{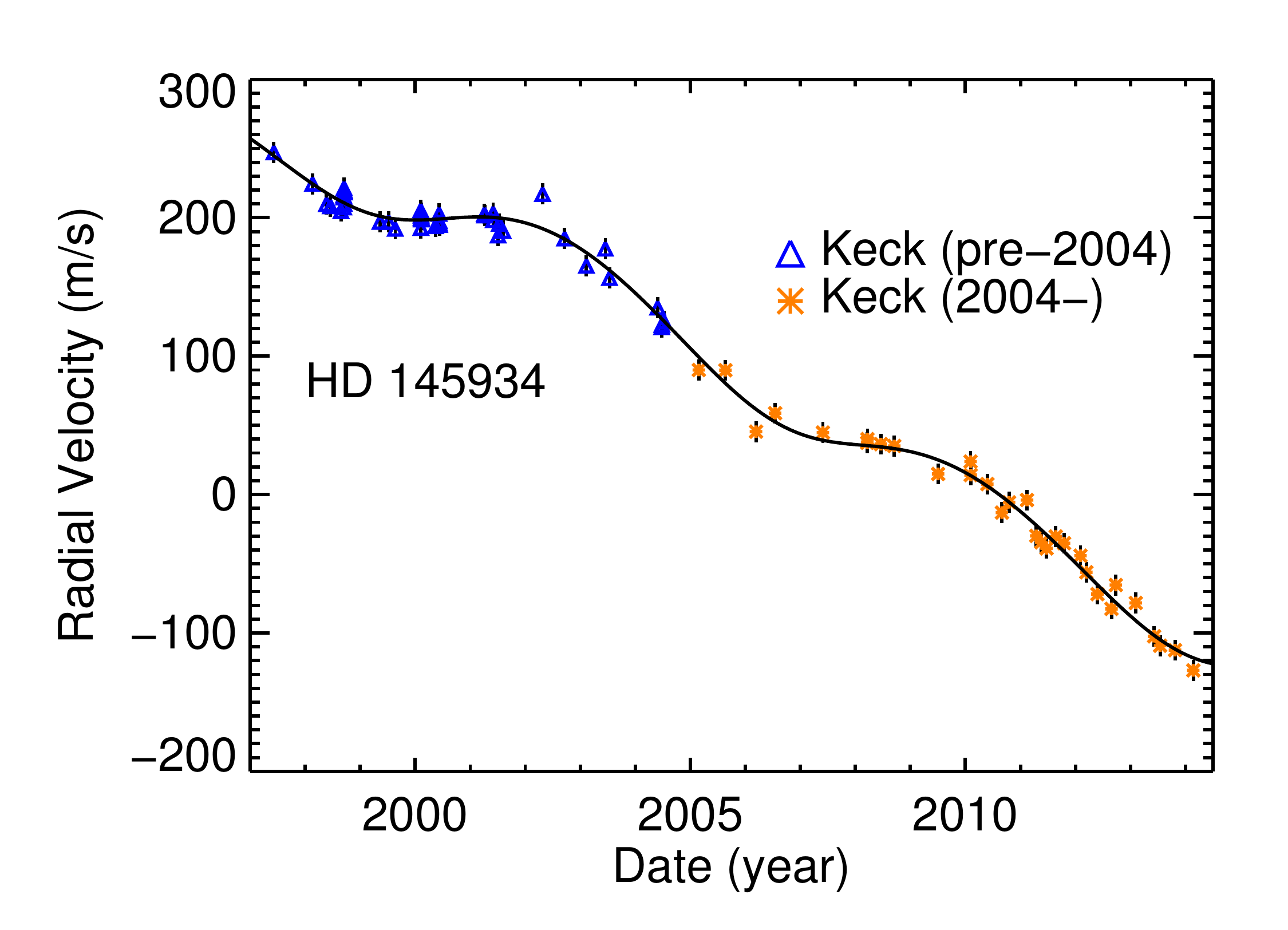}
    \label{145934sys}}
\quad
\subfigure[Residuals]{
    \includegraphics[scale=0.32]{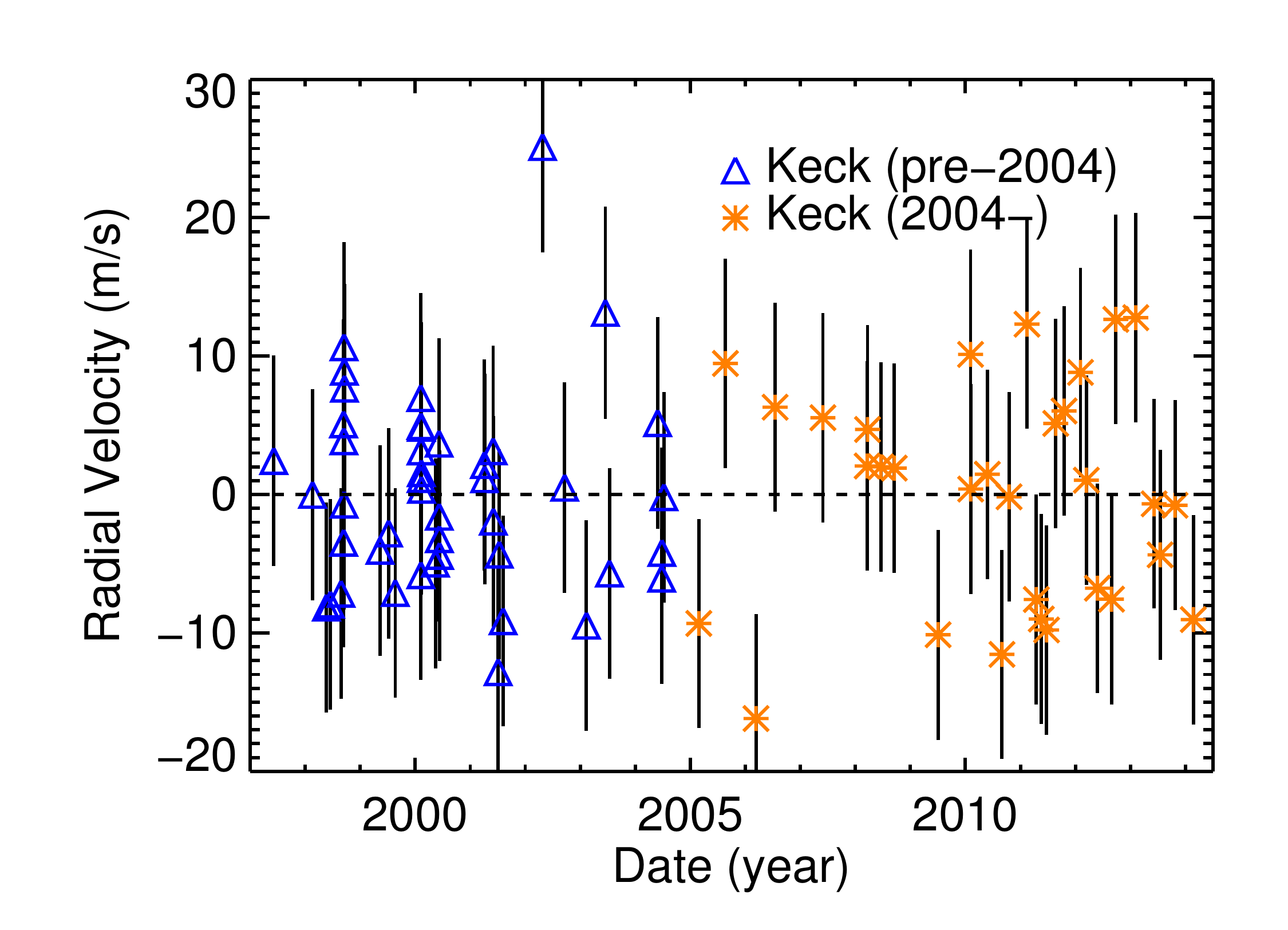}
    \label{145934res}}
\quad
\subfigure[HD 145934 \textit{b}]{
    \includegraphics[scale=0.32]{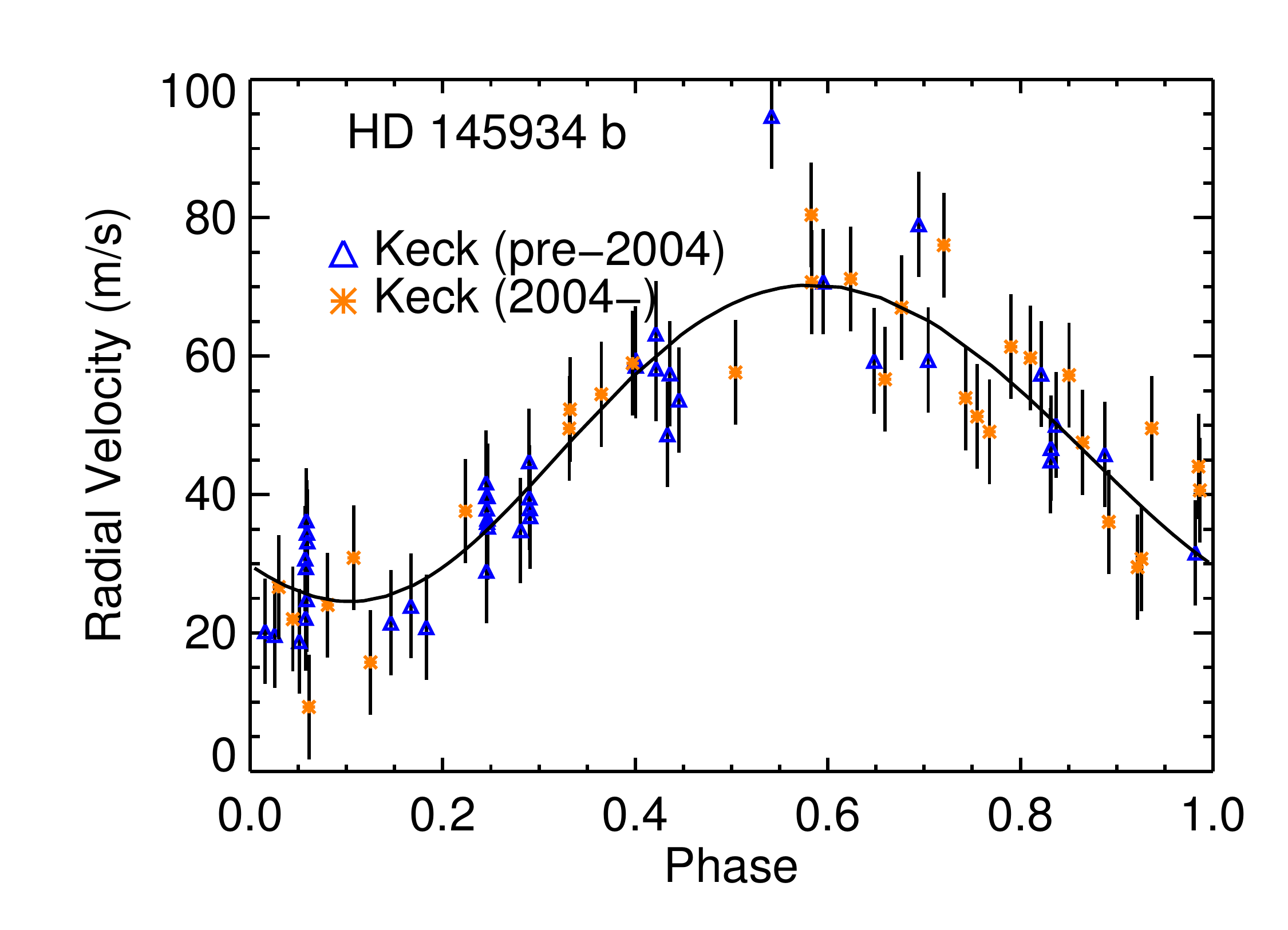}
    \label{145934b}}
\quad
\subfigure[HD 145934 outer companion]{
    \includegraphics[scale=0.32]{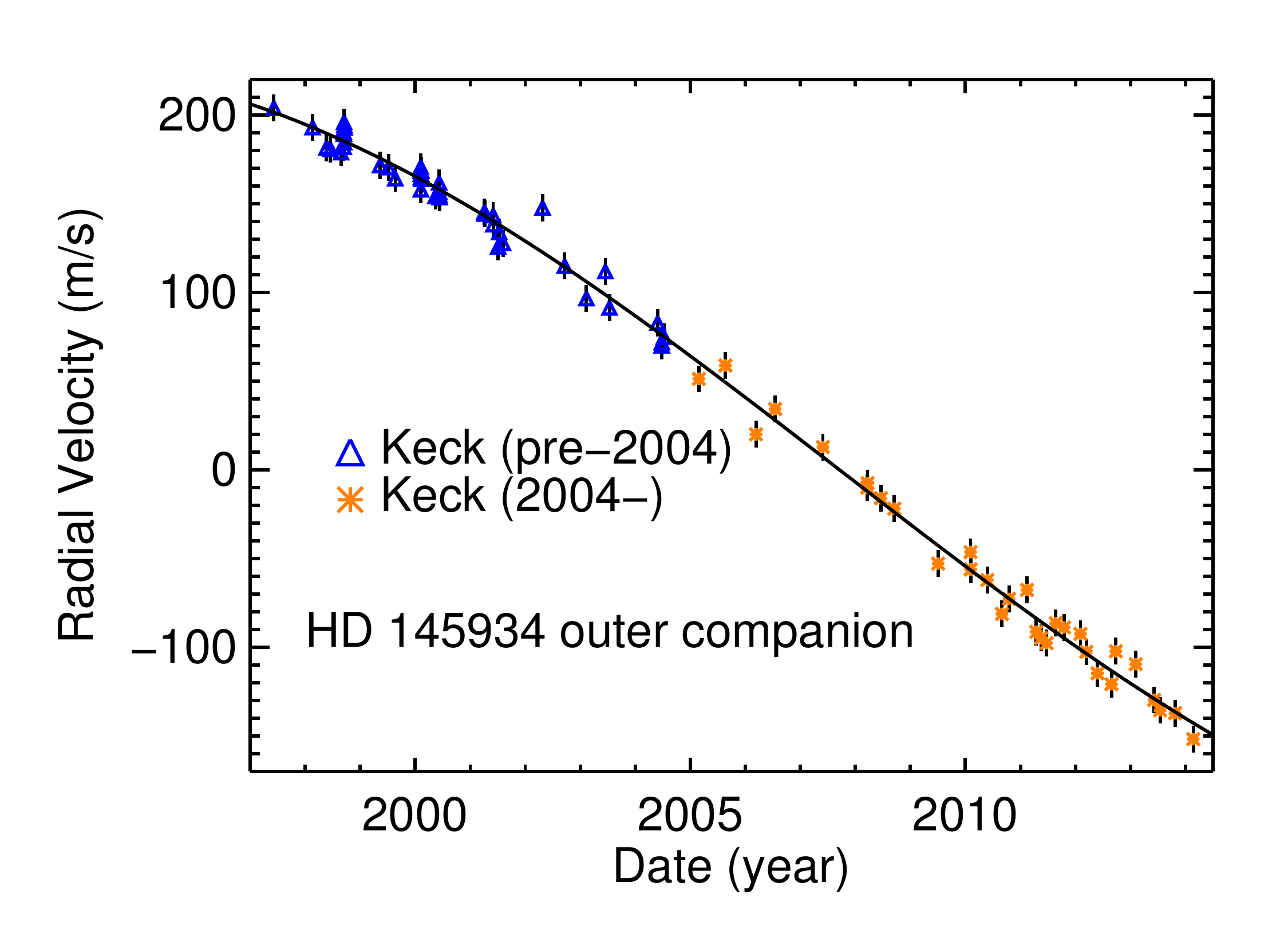}
    \label{145934c}}
\caption[Radial velocity and Keplerian fit for HD 145934 \textit{b}.]{Radial velocity and Keplerian fit for HD 145934. Solid lines represent the best-fit Keplerian orbit. 

\ref{145934sys} Keck RVs overplotted by best-fit two-planet Keplerian model that includes a long-period companion to account for curvature. \ref{145934res} Residuals of the RVs with the best-fit two-planet Keplerian model subtracted. \ref{145934b} and \ref{145934c}: the RV curves for HD 145934 \textit{b} and the outer long-period companion, respectively.}
\label{145934plot}
\end{figure}

In our analysis of the 75 HIRES velocities for HD 145934, we note the slight overall curvature present (see Figure \ref{145934c}).  To account for the curvature using {\tt RVLIN}, which (at the moment) only accommodates purely linear trends, we treated HD 145934 as a two-companion system, with the outer companion having a very long (60 yr) orbital period and circular orbit.  There is not enough information in our time series for the resulting orbital parameters of the outer companion to be meaningful, but this approach provides us sufficient flexibility to fit out the low-frequency power contributed by the binary companion.  Equation~(\ref{mm}) constrains the minimum mass of the companion to be at least $21 M_{\rm Jup}$.

To determine the effects of modeling the ostensible stellar companion with our choice of orbital parameters on the planet's parameters, we checked first the impact of letting eccentricity be a free parameter. The best-fit eccentricity is close to circular (~0.05), so our choice of fixed $e=0$ is not strongly affecting our analysis.  We also changed the (fixed) period of the stellar companion to take values between $50$ and $80$ yr (guesses outside the range returned poor fits, but given the nonlinear nature of the problem this does not necessarily reflect an actual upper limit to the companion's period).  We found that the choice of period did not have significant impact on the parameters of the planet. For example, the best-fit values for the period of HD 145934 \textit{b} varied on the order of 10 days for different outer companion periods. The minimum mass varied on the order of 0.1 $M_{\rm Jup}$. These differences are all well within 1$\sigma$ of our presented set of parameters.  We conclude that our modeling of the outer companion is sufficiently flexible to have no important effects on our estimates of the planet's orbital parameters.

Given that the rms of the residuals to the fit without stellar jitter is 7.83 m s$^{-1}$, we assume a stellar jitter of 7.5 m s$^{-1}$ in our fit. \citet{hekker2006} performed a survey of stable K giants with jitters lower than 20 m s$^{-1}$. The most stable of that sample range between 6 and 15 m s$^{-1}$, so our choice of jitter is reasonable and also consistent with the residuals. The residuals to the resulting best-fit Keplerian model have rms of 7.80 m s$^{-1}$ and \kain of 1.05. We find that HD 145934 \textit{b} has a period of 7.48 $\pm$ 0.27 yr, an orbital eccentricity of 0.053$_{-0.063}^{+0.053}$ , and a semi-amplitude of $22.9 \pm 2.6$ m s$^{-1}$. The minimum mass of the planetary companion is 2.28 $\pm$ 0.26 $M_{\rm Jup}$.

The presence of curvature in the binary companion's orbit implies that either it is highly eccentric and near periapse, or that we have observed a nonnegligible portion of its orbit.  The latter is more likely, and implies that its orbital period is a several or dozens of decades, not millennia.  

\section{Discussion} \label{discussion}

Our analysis of 13 exoplanets uses recent Keck-HIRES radial velocities and other published data. We see that there is need for follow-up work, as in the cases of GJ 849 and HD 145934 for better constraints and further analysis. In the instance of HD 66428, whose residuals show a previously unseen linear trend, we will monitor for the completion of orbits or to rule the companion out as a planet.

We have reduced the uncertainties in the parameters for many planets. The up-to-date HIRES data allowed us to place upper limits or constrain several orbits. From our sample, we identify two planets as Jupiter analogs around Sun-like due to similarities in semimajor axis (5.2 AU): HD 24040 \textit{b} and HD 187123 \textit{c}, although both are much more massive than Jupiter, and the latter's orbit is somewhat eccentric. We have discovered a new planet, HD 145934 \textit{b}, and its host star's residuals show curvature whose velocity semiamplitude is indicative of a probable stellar or brown dwarf companion. 

We confirm GJ 849 \textit{c}, and find that it is the planet with the longest known period around an M dwarf so far.  GJ 849 is a rare system in that it is a multi-giant-planet system around an M-dwarf.  In all of our multi-planet systems, the inner planet is less massive, though this fact is certainly influenced by the soft decrease in semiamplitude with orbital distance ($K\propto a^{
-1/2}$).  HD 66428 may be a case where the planet's high eccentricity and the presence of a linear trend in the system are signs the outer companion has affected the inner planet's orbit, as \citet{kane2014} found in the case of HD 4203.
 
All of these systems, but perhaps especially the ``highly hierarchical'' systems \citep{wright2010} HD 187123 and HD 217107, will be valuable for reconciling observations and the theory of planetary migration.   These two systems are at present the only known examples of systems containing a hot Jupiter (gas giant with $P<10$ days and $M \sin i > 0.1 M_{\rm Jup}$) and a very-long-period planet ($P > 5$ yr) with a well determined orbit.  In both cases, the outer planet is $\sim 3$ times the mass of the inner planet, and there is no evidence of other planets in the system.   
 
There are only two other systems with hot Jupiters and well-constrained long-period ($P > 1$ yr) outer planets: HIP 14810 \citep{butler2006,wright14810} and HAT-P-13 \citep{bakos09,winn10}.  The former case remains anomalous in that the innermost planet is the most massive, with $M \sin i = 3.9 M_{\rm Jup}$ (the outermost planet has $M \sin i = 0.6 M_{\rm Jup}$ and $P = 2.6$ yr; there is also a third, intermediate planet in the system).  The latter case has an especially high mass ratio, having a highly eccentric $M \sin i > 14  M_{\rm Jup}$ outer planet and an inner, transiting planet with $M = 0.86  M_{\rm Jup}$.   We know from both RV studies \citep{2009wright} and the {\it Kepler} results \citep{Latham2011} that ``hot Jupiters are lonely'', at least when it comes to companions within $\sim 1$ AU.  Continued long-term monitoring of other hot Jupiters will establish whether they have frequently have ``cold friends'' at larger orbital distances \citep[e.g.,][]{Knutson14}.

\acknowledgments

We thank the many observers who contributed to the Lick and Keck-HIRES measurements
reported here, especially John Johnson, Debra Fischer, Steven Vogt and R.\ Paul Butler.  We gratefully acknowledge the efforts and  dedication of the Keck Observatory staff, especially Scott Dahm, Hien Tran,
  Grant Hill, and Gregg Doppmann for support of HIRES and Greg Wirth for support of remote
  observing.   

We thank NASA, the University of California, and the University of Hawaii for their allocations of time on the Keck I telescope.  Data presented herein were obtained at the W.\ M.\ Keck Observatory from telescope time allocated to the National Aeronautics and Space Administration through the agency’s scientific partnership with the California Institute of Technology and the University of California. The Observatory was made possible by the generous financial support of the W.\ M.\ Keck Foundation.  We wish to recognize and acknowledge the very significant cultural role and reverence that the summit of Mauna Kea has always had within the indigenous Hawaiian community.  We are most fortunate to have the opportunity to conduct observations from this mountain.  
 
We thank the many astronomers that contributed to the published CORALIE, ELODIE, SOPHIE, and HARPS RV measurements of these important long-period systems.  In particular, we thank Xavier Bonfils for providing us with the HARPS radial velocity data of GJ 849 used in \citet{bonfils2013}. 

The authors acknowledge the Pennsylvania State Research Computing and Cyberinfrastructure Group for providing computational resources and support that have contributed to the results reported within this paper

This work was partially supported by: NASA Keck PI Data Awards, administered by the NASA Exoplanet Science Institute, including awards 2007B\_N095Hr, 2010A\_N147Hr, 2011A\&B\_N141Hr, \& 2012A\_N129Hr; NASA Origins of Solar Systems grant NNX09AB35G; NASA Astrobiology Institute grant NNA09DA76A; and the Center for Exoplanets and Habitable Worlds (which is supported by the Pennsylvania State University, the Eberly College of Science, and the Pennsylvania Space Grant Consortium).  We acknowledge NSF grant AST-1211441. 

This work has made use of data from the SIMBAD Astronomical Database (operated at CDS, Strasbourg, France); NASA's Astrophysics Data System Bibliographic
Services; and of the Exoplanet Orbit Database and Exoplanet Data Explorer at exoplanets.org.

Facility: \facility{Keck: I}

%%%%%%%%RV/Offsets table%%%%%%%
\clearpage
\input{offsets_table.tex}
\clearpage
\newpage
\clearpage
%\begin{landscape}
\begin{turnpage}
%%%%%%%% Parameters %%%%%%%%%
\begin{deluxetable}{lllcrrrrrrrrr}
\tablecaption{List of Stellar Parameters \label{stellar}}
\tablewidth{0pt}
\tabletypesize{\scriptsize}
\tablehead{
\colhead{Name} & \colhead{R.A.} & \colhead{Decl.}&\colhead{Sp. Ty.} & \colhead{$V$} & \colhead{$B-V$} & \colhead{Dist} & \colhead{$T_{\rm eff}$} & \colhead{$\log(g)$} & \colhead{[Fe/H]} & \colhead{$v\sin(i)$ } & \colhead{$M_{\star}$} & \colhead{$R_{\star}$}\\
& & & & &  &(pc) &(K) & (cgs) & (dex) & (km s$^{-1}$)& ($M_{\odot}$) & ($R_{\odot}$)\\
 (1) & (2) & (3) & (4) & (5) &(6) & (7) & (8) & (9) & (10) & (11) & (12) &(13)
}
\startdata
HD 24040 &03 50 22.9 &+17 28 34.9 &G0 & 7.5&0.65 &46.6(1.6)&5853(44)&4.361(60)&0.206(30)&2.39(50)&1.18(10)&1.154(39) \\
HD 66428 &08 03 28.7 &$-$01 09 45.7 &G5 & 8.3&0.71 &54.9(3.2)&5752(44)&4.490(60)&0.310(30)&0.00(50)&1.061(63)&0.980(34) \\
HD 74156 &08 42 25.1 &+04 34 41.1 &G0 & 7.6&0.58 &64.4(2.2)&6068(44)&4.259(60)&0.131(30)&4.32(50)&1.238(42)&1.345(44) \\
HD 145934&16 13 09.9 &+13 14 22.1 & K0 & 8.5 & 1.05& \nodata & \nodata& 3.23(6) & \nodata& \nodata  &1.748(105) & 5.38(44)\\
 HD 183263 &19 28 24.6 &+08 21 28.9 &G2IV & 7.9&0.68 &55.1(2.8)&5936(44)&4.403(60)&0.302(30)&1.56(50)&1.121(52)&1.117(38) \\
HD 187123 &19 46 58.1 &+34 25 10.3 &G5V & 7.8&0.66 &48.3(1.2)&5815(44)&4.359(60)&0.121(30)&2.15(50)&1.037(25)&1.143(39) \\
HD 217107 & 22 58 15.5&$-$02 23 43.4 &G8 & 6.2&0.74 &19.86(15)&5704(44)&4.541(60)&0.389(30)&0.00(50)&1.108(43)&1.500(30) \\
GJ 849 &22 09 40.3 &$-$04 38 26.6 &M3.5V & 10.4&1.5 &9.10(17) &3601(19) \tablenotemark{a} & \nodata &0.31(12) \tablenotemark{a}& \nodata &0.490(49) \tablenotemark{a} & \nodata \\
\enddata
\tablecomments{We use parenthetical notation for the uncertainties to display the data in a succinct manner. The least significant digit of the uncertainty, shown in parentheses, has the same place value as that of the quantity. For example, ``0.460(30)'' is equivalent of ``0.046 $\pm$ 0.030'', and ``5898(44)'' is equivalent of ``5898 $\pm$ 44''. Unless stated otherwise: the values in Columns 2--6 are from various sources collated in the SIMBAD Astronomical Database; distances are from \citet{hip}; data in Columns 8--11 are from \citet{valenti2005}; data in Column 12 are from \citet{takeda2007}; and we calculated the stellar radii based on \citet{torres2010}.}
\tablenotetext{a}{For GJ 849, $T_{\rm eff}$ is from \citet{rojas2012}, [Fe/H] is from \citet{terrien2012}, while $M_{\star}$ is from \citet{montet2014}.}
\end{deluxetable}
\clearpage 
%\end{landscape}
\end{turnpage}

%\clearpage 
%\begin{landscape}
\begin{turnpage}
\begin{deluxetable}{rlrp{0pt}lrp{0pt}lrp{0pt}lrp{0pt}lrp{0pt}lrp{0pt}lrp{0pt}l}
\tablecaption{List of Keplerian Orbital Parameters \label{orb}}
\tablewidth{8in}
%\rotate
\tabletypesize{\scriptsize}
\tablehead{
\multicolumn{2}{c}{{Name}} & \multicolumn{3}{c}{{$P$}} & \multicolumn{3}{c}{{$T_{\rm p}$}} & \multicolumn{3}{c}{{$e$}} & \multicolumn{3}{c}{{$\omega$}} & \multicolumn{3}{c}{{$K$} }& \multicolumn{3}{c}{{$M \sin{i}$}} & \multicolumn{3}{c}{{$a$}} \\
 & &\multicolumn{3}{c}{(days)} & \multicolumn{3}{c}{(BJD -- 2440000)} &&& & \multicolumn{3}{c}{($^{\circ}$)} & \multicolumn{3}{c}{(m s$^{-1}$)} & \multicolumn{3}{c}{($M_{\rm Jup}$)} & \multicolumn{3}{c}{(AU)}
}
\startdata
HD 24040 & \textit{b}\tablenotemark{a} & 3490 & $\pm$ &  25 & 16670 & $\pm$ &  240 & 0.047 & $\pm$ &  0.020 & 67 & $\pm$ &  24 & 51.8 & $\pm$ &  1.6&4.10 & $\pm$ &  0.12&4.637 & $\pm$ &  0.067 \\
HD 66428 & \textit{b}\tablenotemark{b} & 2293.9 &$\pm$& 6.4 & 12278 &$\pm$& 16 & 0.440 &$\pm$& 0.013 & 180.4 &$\pm$& 2.6 & 52.6 &$\pm$& 1.1&3.194 &$\pm$& 0.060&3.471 &$\pm$& 0.069 \\
 HD 74156 & \textit{b} &51.6385 & $\pm$ &  0.0015 & 10793.39 & $\pm$ &  0.11 & 0.6380 & $\pm$ &  0.0061 & 175.35 & $\pm$ &  0.92 & 109.1 & $\pm$ &  1.6&1.778 & $\pm$ &  0.020&0.2916 & $\pm$ &  0.0033 \\ 
 & \textit{c} & 2448.9 & $\pm$ &  5.5 & 8559 & $\pm$ &  15 & 0.3829 & $\pm$ &  0.0080 & 268.9 & $\pm$ &  1.6 & 112.6 & $\pm$ &  1.3&7.997 & $\pm$ &  0.095&3.820 & $\pm$ &  0.044 \\
HD 145934 & \textit{b} &2730 &$\pm$& 100 & 11430 &$\pm$& 370 & $0.053$&$^{+}_{-}$&$_{0.063}^{0.053}$ & 215 &$\pm$& 62 & 22.9 &$\pm$& 2.6&2.28&$\pm$& 0.26&4.60 &$\pm$& 0.14 \\
HD 183263 & \textit{b} &625.10 &$\pm$& 0.34 & 12113.0 &$\pm$& 2.4 & 0.3728 &$\pm$& 0.0065 & 232.9 &$\pm$& 1.4 & 86.16 &$\pm$& 0.79&3.635 &$\pm$& 0.034&1.486 &$\pm$& 0.023 \\
 & \textit{c}&4684 &$\pm$& 71 & 10430 &$\pm$& 310 & 0.051 &$\pm$& 0.010 & 299 &$\pm$& 22 & 77.5 &$\pm$& 1.1&6.90 &$\pm$& 0.12&5.69 &$\pm$& 0.11 \\
HD 187123 & \textit{b} & 3.0965886 &$\pm$& 0.0000043 & 14342.87 &$\pm$& 0.30 & 0.0093 &$\pm$& 0.0046 & 360 &$\pm$& 200 & 68.91 &$\pm$& 0.36&0.5074 &$\pm$& 0.0026&0.04213 &$\pm$& 0.00034 \\
 & \textit{c}  & 3324 &$\pm$& 46 & 13625 &$\pm$& 40 & 0.280 &$\pm$& 0.022 & 258.5 &$\pm$& 3.9 & 25.10 &$\pm$& 0.44&1.818 &$\pm$& 0.035&4.417 &$\pm$& 0.054 \\
 HD 217107 & \textit{b} & 7.126846 &$\pm$& 0.000013 & 14395.789 &$\pm$& 0.025 & 0.1283 &$\pm$& 0.0027 & 24.0 &$\pm$& 1.3 & 140.30 &$\pm$& 0.40&1.4135 &$\pm$& 0.0042&0.07505 &$\pm$& 0.00097 \\
  & \textit{c} & 5189 &$\pm$& 21 & 10770 &$\pm$& 16 & 0.3848 &$\pm$& 0.0086 & 206.3 &$\pm$& 1.7 & 53.41 &$\pm$& 0.75&4.513 &$\pm$& 0.072&6.074 &$\pm$& 0.080\\
GJ 849 & \textit{b}\tablenotemark{c} & 1924 &$\pm$& 15 & 13770 &$\pm$& 150 & 0.038 &$\pm$& 0.019 & 66 &$\pm$& 28 & 23.96 &$\pm$& 0.94 & 0.911 &$\pm$& 0.036 & 2.39 &$\pm$& 0.082\\
 & \textit{c}\tablenotemark{c} & 5520 &$\pm$& 390\tablenotemark{d} & 14320 &$\pm$& 690 & 0.087 &$\pm$& 0.056 & 172 &$\pm$& 50 & 17.5 &$\pm$& 1.1 & 0.944 &$\pm$& 0.070 & 4.82 &$\pm$& 0.21
\enddata
\tablenotetext{a}{The fit includes a linear trend of $1.8 \pm 0.4$ m s$^{-1}$ yr$^{-1}$}
\tablenotetext{b}{The fit includes a linear trend of $-3.4$ $\pm$ 0.2 m s$^{-1}$ yr$^{-1}$}
\tablenotetext{c}{Except where noted, these parameter uncertainties were computed using only the stable bootstrapping realizations.}
\tablenotetext{d}{This parameter uncertainty was computed using the \kai\ map.}
\end{deluxetable}
\clearpage
%\end{landscape}
\end{turnpage}

%%%%%%%%%%%%%%   RV TABLES  %%%%%%%%%%%%%%
%\LongTables
\input{sample.tex}

%\input{24040rv.tex}
%\input{66428rv.tex}
%\input{187123rv.tex}
%\input{74156rv.tex}
%\input{183263rv.tex}
%\input{217107rv.tex}
%\input{gj849rv.tex}
%\input{145934rv.tex}
%%%%%%%%%%%%%%%%%%%%%%%%%%%%%%%%%%%%%%%%%%
\newpage

\end{document}

%% file: offsets_table.tex
%\documentclass[manuscript]{aastex}

%\begin{document}
\begin{deluxetable}{ccccccc}
\tablecaption{Summary of Radial Velocity Data (Number of Observations) and Mean Uncertainties \label{rv_sum}}
\tablewidth{0pt}
\tabletypesize{\scriptsize}
\tablehead{
\colhead{} &\colhead{Instrument} & \colhead{Mean Unc. (m s$^{-1}$)} & \colhead{$N_{\rm obs}$} & \colhead{Span (yr)} & \colhead{$N_{\rm new}$} & \colhead{Offset from Instrument 1 (m s$^{-1}$)}
}
\startdata
\multicolumn{7}{c}{HD 24040*} \\
\hline
 1 & SOPHIE&4.27 & 13&2008--2010 & \nodata & \nodata\\
  2  &ELODIE &11.48 &47 &1997--2005 &\nodata & $-53.19\pm4.37$ \\
  3 & HIRES (pre-upgrade) & 1.35 & 20& 1998--2004 & \nodata & $-44.73\pm3.80$\\
  4 & HIRES (post-upgrade) & 1.56 & 27& 2004--late 2013& 22 & $-34.28\pm1.69$\\
  \hline\hline
 \multicolumn{7}{c}{HD 66428. Jitter = 3 m s$^{-1}$} \\
 \hline
1 & HIRES (pre-upgrade)& 1.22&22 &2000--2004 &\nodata & \nodata\\
    2   & HIRES (post-upgrade)& 1.05&33 &2004--late 2013 & 26 &$2.8\pm1.8$\\
       \hline \hline
  \multicolumn{7}{c}{HD 74156*} \\
  \hline
 1  &CORALIE &8.52 & 44& 2001--2003& \nodata &\nodata\\
  2   & HRS& 8.34& 82&2004--2007 &\nodata &$-32.5\pm3.07$  \\
  3  &ELODIE &12.74 & 51&1998--2003& \nodata &$32.16\pm2.82$  \\
  4  &HIRES (pre-upgrade) &1.99 &9 &2001--2004 & \nodata &$47.02\pm2.6$ \\
  5  & HIRES (post-upgrade)&2.87 & 43& 2004--late 2013& 31 & $66.73\pm2.6$\\
     \hline \hline
     \multicolumn{7}{c}{HD 145934. Jitter = 7.5 m s$^{-1}$} \\
  \hline
1   &HIRES (pre-upgrade) &1.22 &44 & 1997--2004& \nodata& \nodata\\
 2  & HIRES (post-upgrade)& 1.00& 31& 2004--early 2014& 75 & $14.99\pm9.70$\\
   \hline \hline
     \multicolumn{7}{c}{HD 183263. Jitter = 3.2 m s$^{-1}$} \\
  \hline
1   & HIRES (pre-upgrade)&1.6 & 31& 2001--2004& \nodata & \nodata\\
 2  &HIRES (post-upgrade)&1.23 & 11& 2004--mid-2013 & 24 &   $4.04\pm4.4$ \\
   \hline \hline
     \multicolumn{7}{c}{HD 187123. Jitter = 2.23 m s$^{-1}$} \\
  \hline
1   & HIRES (pre-upgrade)& 1.22&64 &1997--2004 & \nodata & \nodata\\
 2  &HIRES (post-upgrade) &1.19 & 46&2004--mid-2013 & 40 &$-1.58\pm0.88$ \\
   \hline \hline
     \multicolumn{7}{c}{HD 217107*} \\
  \hline
  1 & Hamilton&4.709 &121 &1998--2007 & \nodata & \nodata\\
2   & CORALIE& 9.175&63 &1998-1999 &\nodata & $-402.3\pm1.74$\\
3   & HIRES (pre-upgrade)& 1.414&63 &1998--2004 & \nodata & $17.6\pm1.38$\\
4   & HIRES (post-upgrade)&1.022 &68&2004--late 2013  & 31 & $25.68\pm1.87$\\
   \hline \hline
     \multicolumn{7}{c}{GJ 849. Jitter = 3  m s$^{-1}$} \\
  \hline
1   &HIRES (pre-upgrade) &3.19 & 24&1997--2004 & \nodata & \nodata\\
2   &HIRES (post-upgrade) &1.48& 58& 2004--late 2013& 3 & $-4.24\pm2.24$\\
   3   & HARPS&1.05&35 &2003--2008 &\nodata & $18.36\pm2.1$
  % & & & & & \\
   % & & & & & \\

\enddata
\tablecomments{For each star, we list the chosen jitter, instruments, the mean uncertainty corresponding to the data from each instrument before jitter is applied, number of observations, date range of observations, the number of new points this work has added, and offsets from instrument 1. In a few cases, marked by an asterisk ($^*$), we add jitter instrument-by-instrument. We split data from HIRES by the 2004 upgrade.\\
For references of data, see text. }
\end{deluxetable} 

%\end{document}

%% file: sample.tex
\begin{deluxetable}{cccc}
\tablecaption{Radial Velocities Measured for HD 24040 \label{24040rv}}
\tablewidth{0pt}
\tabletypesize{\scriptsize}
\tablehead{
\colhead{BJD} & \colhead{RV} & \colhead{$\pm 1 \sigma$} & \colhead{Tel} \\
--2440000 & (m s$^{-1}$)& (m s$^{-1}$) &
}
\startdata
 10838.77321 & $-95.2$ & 1.5 & HIRES \\
11043.11965 & $-85.4$ & 1.5 & HIRES \\
11072.03904 & $-87.0$ & 1.5 & HIRES \\
11073.00232 & $-87.8$ & 1.1 & HIRES
\enddata
\tablecomments{This table is available in its entirety in machine-readable form.
}
\end{deluxetable}